\documentclass[aps,prd,twocolumn,nofootinbib,superscriptaddress]{revtex4-2}

\usepackage[table]{xcolor}
\usepackage{subcaption}

\newcommand{\corrcell}[1]{%
  \ifdim #1 pt < 0pt
    \cellcolor{red!\numexpr-100*#1\relax!white}%
  \else
    \cellcolor{blue!\numexpr100*#1\relax!white}%
  \fi
  #1
}

\usepackage{amsmath,amssymb}
\usepackage{graphicx}
\usepackage{xcolor}
\usepackage[colorlinks=true,linkcolor=blue,citecolor=blue,urlcolor=blue]{hyperref}
\usepackage{siunitx}

\usepackage[normalem]{ulem}

\begin{document}

\title{Calibrated correlation between heavy-quark masses \\ and Hadronic Vacuum Polarization observables at the precision frontier}

\author{Arnau Beltran}
\affiliation{%
PRISMA++ Cluster of Excellence and Institut f\"ur Kernphysik, Johannes Gutenberg-Universit\"at
Mainz, 55099 Mainz, Germany.
}%
\author{Pere Masjuan}
\affiliation{%
Grup de F\'{i}sica Te\`orica, Departament de F\'{i}sica, Universitat Aut\`onoma de Barcelona, Institut de F\'{i}sica d'Altes Energies (IFAE),
and The Barcelona Institute of Science and Technology (BIST), Campus UAB, E-08193 Bellaterra (Barcelona), Spain. 
}%
\author{Antonio Rivera}
\affiliation{%
Grup de F\'{i}sica Te\`orica, Departament de F\'{i}sica, Universitat Aut\`onoma de Barcelona, Institut de F\'{i}sica d'Altes Energies (IFAE),
and The Barcelona Institute of Science and Technology (BIST), Campus UAB, E-08193 Bellaterra (Barcelona), Spain. 
}%

\date{\today}
\begin{abstract}
The theoretical prediction of the muon anomalous magnetic moment $a_\mu$ depends crucially on the Hadronic Vacuum Polarization (HVP), and the tension between its dispersive and lattice-QCD determinations remains unresolved. We show that part of this puzzle can be addressed in the heavy-quark sector, where both descriptions are theoretically clean, by recognizing that the heavy-quark mass and its contribution to $a_\mu$ are not independent quantities: both follow from integrals of the same hadronic spectral function, differing only in their integration kernel. Promoting this kernel to a free choice within the relativistic QCD Sum Rules used to determine heavy-quark masses, we break with the conventional notion of a single valid sum rule and instead determine the mass and its HVP contribution simultaneously, from a common, self-consistent framework. This intrinsic construction exploits the anticorrelation between the two quantities to sharpen the final uncertainty, and turns the residual disagreement between the perturbative and hadronic descriptions of the observable into a direct observable-specific diagnostic of residual theory/model dependence, including duality-violation and continuum-modeling effects, unavailable to a determination of the mass alone. We obtain $a_\mu^{\rm HVP_{c+b},LO} =(14.46(13)+0.3009(17))\times 10^{-10}$ at leading and $a_\mu^{\rm HVP_{c+b}, NLO_{a,b}} = ( -0.5738(95) - 0.01822(13) )\times 10^{-10}$ at next-to-leading order, for charm and bottom contributions, respectively. We compare our next-to-leading-order results with its first available lattice determination, finding good agreement in the charm sector. As a byproduct, we obtain $\hat m_c(\hat m_c)=1267.1(6.8)$ MeV and $\hat m_b(\hat m_b) = 4182.3(7.2)$ MeV, with unprecedented phenomenological precision. Beyond these results, the construction introduced here defines a general strategy, applicable to any observable expressible as a kernel-weighted dispersive integral, for extracting correlated hadronic quantities from a single, unified sum rule.
\end{abstract}

\maketitle

\section{Introduction}
\label{sec:intro}

The muon anomalous magnetic moment, $a_\mu$, is one of the most precisely measured and most precisely predicted quantities in particle physics, and the comparison between the two continues to serve as a sensitive test of the Standard Model~\cite{Jegerlehner:2017gek,Aoyama:2020ynm,Aliberti:2025beg}. The dominant source of theoretical uncertainty in the Standard Model prediction is the Hadronic Vacuum Polarization (HVP) contribution, which cannot be computed in perturbation theory at the energies relevant to the muon and must instead be obtained either from a dispersive analysis of $e^+e^-\to{\rm hadrons}$ data or from lattice QCD. These two approaches probe the same underlying physics through different representations, dispersive in the timelike region for the former, perturbative in the spacelike region for the latter, and their comparison is therefore intrinsically nonlocal. Global determinations of $a_\mu^{\rm HVP}$ obtained from these two approaches currently show differences whose origin is not yet fully understood~\cite{Aoyama:2020ynm,Aliberti:2025beg}, and identifying which sources of uncertainty are responsible, and to what extent, remains an open problem.

Part of the difficulty is that most of the HVP contribution comes from the light-quark sector, where the interplay of resonances, thresholds, and the transition to a perturbative description is not easily disentangled, and neither the dispersive nor the lattice determination admits a fully independent cross-check in this regime. The heavy-quark sector offers a more tractable setting to make progress. The charm and bottom contributions to $a_\mu^{\rm HVP}$ are numerically smaller, but both the timelike and spacelike descriptions of the corresponding vector-current correlator are under good perturbative control near threshold, and the same correlator that determines the HVP contribution also determines the heavy-quark mass through the well-established framework of moment sum rules~\cite{Novikov:1977dq,Shifman:1978bx,Shifman:1978by}. This makes the heavy-quark sector a theoretically clean environment in which to test, quantitatively, how much of the tension between dispersive and lattice HVP determinations can be attributed to sources of uncertainty that have not yet been systematically explored.

Previous applications of the moment sum rule method have used it either to determine the heavy-quark mass, treating $a_\mu^{q}$ as a separate quantity to be computed afterward with the extracted mass as external input, or to compute $a_\mu^{q}$ directly from a model of the spectral function without reference to the mass determination. In this work we observe that these are not, in fact, independent problems. The heavy-quark mass and its HVP contribution are both weighted integrals of the same hadronic spectral function, differing only in the choice of integration kernel, so any determination of one necessarily carries information about the other. We exploit this structural connection directly, rather than treating it as a correlation to be estimated after the fact, by promoting the HVP kernel itself to the weight of a generalized moment sum rule. This allows $\hat m_q$ and $a_\mu^{q}$ to be extracted simultaneously from a single, self-consistent set of sum rules, with their correlation built into the construction from the outset.

This generalization has two direct benefits. First, since the anticorrelation between $\hat m_q$ and $a_\mu^{q}$ is intrinsic to the dispersive framework, accounting for it consistently reduces the final uncertainty on $a_\mu^{q}$ relative to a determination that treats the mass as an independent external input. Second, because the generalized sum rule enforces agreement between the perturbative and hadronic descriptions of $a_\mu^{q}$ directly, rather than of an auxiliary moment only indirectly related to it, the residual spread between these two descriptions at moment pairs other than the one used for the extraction becomes a direct, observable-specific diagnostic of residual theory/model dependence, including duality-violation and continuum-modeling systematics, one that is not accessible from the mass determination alone.

Using this framework, we determine $\hat m_c$, $\hat m_b$, and the corresponding LO and NLO contributions to $a_\mu^{\rm HVP_{c,b}}$, extending the existing moment sum-rule treatment of both quark sectors~\cite{Erler:2002bu,Erler:2016atg,Erler:2022mzd} with an updated continuum ansatz and, for charm, an extended set of explicit narrow resonances. We compare our LO results with existing dispersive and lattice determinations, and our NLO results with the first lattice calculation available at this order, finding agreement in both cases but only after correlations have been taken into account.

The remainder of this paper is organized as follows. Section~\ref{sec:hvp} reviews the dispersive representation of $a_\mu^{\rm HVP}$ and the local-duality assumption it requires. Section~\ref{sec:sumrules} presents the moment sum-rule formalism for the heavy-quark mass, including the two extensions to the continuum ansatz adopted in this work. Section~\ref{sec:generalized} introduces the new generalized, kernel-weighted, correlated sum rules that allow $\hat m_q$ and $a_\mu^{q}$ to be determined simultaneously with correlated uncertainty and self-calibration constraint. Section~\ref{sec:results} presents our final results for the charm and bottom sectors, including a calibration through experimental data, and an outlook to beyond leading order operators. Section \ref{sec:comparison} compares them with previous determinations including latest lattice results at NLO. We conclude in Section~\ref{sec:conclusions}.

\section{The hadronic vacuum polarization contribution to $a_\mu$}
\label{sec:hvp}

The leading hadronic contribution to the muon anomalous magnetic moment is given by the dispersive representation \cite{Bouchiat:1961lbg,Brodsky:1967sr,Lautrup:1968tdb,Gourdin:1969dm} 
\begin{equation}
a_\mu^{\text{HVP},(i)} = \left(\frac{\alpha\, m_\mu}{3\pi}\right)^{\!2} \int_{s_{\rm thr}}^{\infty} \frac{ds}{s^2}\, \hat K^{(i)}(s)\, R_{\rm had}(s),
\label{eq:amu_hvp}
\end{equation}
where $\alpha$ is the fine-structure constant, $m_\mu$ the muon mass, and $R_{\rm had}(s) = \sigma(e^+e^- \to \text{hadrons})/\sigma(e^+e^- \to \mu^+\mu^-)$ is the hadronic $R$-ratio. Since the HVP contribution is the only one considered throughout this work, we drop the superscript and write simply $a_\mu^{(i)}$ in what follows. This relation follows from the optical theorem: $R_{\rm had}(s)$ is proportional to the imaginary part of the photon vacuum polarization at timelike momenta, and the Cauchy integral over the cut yields $a_\mu^{\rm HVP}$. The kernel $\hat K^{(i)}(s)$ is a known, sign-definite QED function; the index $i = 2, 4a, 4b$ labels the order in $\alpha$ at which each contribution enters. $\hat K^{(2)}$ is the leading-order (LO) kernel, arising from the one-loop diagram in which the photon propagator is dressed by the hadronic blob. $\hat K^{(4a)}$ and $\hat K^{(4b)}$ are the two next-to-leading-order (NLO) kernels: the former from an additional virtual photon connecting the hadronic blob to the muon line, the latter from light-fermion loop insertions into the photon propagator.

Evaluating Eq.~\eqref{eq:amu_hvp} requires $R_{\rm had}(s)$ over the full positive real axis. Since experimental data only cover a finite energy range, one is forced to switch, at some scale $s_0$, from data to a perturbative-QCD (pQCD) prediction. This switch relies on \emph{local} quark-hadron duality: the assumption that pQCD and $R_{\rm had}(s)$ agree not only on average, but point by point in energy near $\sqrt{s_0}$. The resulting uncertainty is difficult to quantify and is often absent from the quoted error budget of dispersive HVP evaluations.

The heavy-quark sector offers a theoretically clean environment in which to test this assumption in a controlled setting. Both the timelike (dispersive) and spacelike (pQCD) descriptions of the vector-current correlator are available with good perturbative control near the charm and bottom thresholds, and the relevant sum rules only require \emph{global}, not local, duality: the matching is enforced through integrated moments rather than at a fixed energy. This is the framework developed in the following section.

\section{Moment sum rules for the heavy-quark mass}
\label{sec:sumrules}

\subsection{Formalism}

In the following we adopt the formalism presented in Refs.~\cite{Erler:2002bu,Erler:2016atg,Erler:2022mzd}. The HVP function $\Pi_q(q^2)$ is defined through the transverse part of the two-point function of the electromagnetic vector current $j^\mu = Q_q\,\bar q\gamma^\mu q$ of the quark $q$ with electric charge $Q_q$,
\begin{align}
  \Pi_q^{\mu\nu}(q)
  &= i\int d^4x\,e^{iq\cdot x}
    \langle 0|T\{j^\mu(x)j^{\nu\dagger}(0)\}|0\rangle \notag\\
  &= \left(q^\mu q^\nu-g^{\mu\nu}q^2\right)\Pi_q(q^2)\, .
  \label{eq:Pi_def}
\end{align}
$\Pi_q(q^2)$ is analytic in the complex $q^2$ plane with a branch cut along the positive real axis starting at the physical threshold $s_{\rm thr}=4m_q^2$. Its imaginary part along the cut is, by the optical theorem, proportional to the hadronic spectral function,
\begin{equation}
R_q(s) = 12\pi\,\mathrm{Im}\,\Pi_q(s+i\epsilon).
\label{eq:Rq_def}
\end{equation}
Since $\Pi_q(q^2)$ grows logarithmically at large $|q^2|$, it satisfies a once-subtracted dispersion relation \cite{Kniehl:1996rh},
\begin{equation}
\Pi_q(q^2) - \Pi_q(0) = \frac{q^2}{\pi}\int_{4m_q^2}^{\infty} \frac{ds}{s}\,\frac{\mathrm{Im}\,\Pi_q(s)}{s-q^2-i\epsilon}.
\label{eq:disp_rel}
\end{equation}
Taking successive derivatives with respect to $q^2$ at $q^2=0$ generates the tower of moment sum rules
\begin{equation}
\mathcal{M}_n := \frac{12\pi^2}{n!}\frac{d^n}{dt^n}\hat\Pi_q(t)\Big|_{t=0}
= \int_{4\hat m_q^2}^{\infty} \frac{ds}{s^{n+1}}\,R_q(s),
\label{eq:Mn_def}
\end{equation}
for $n\geq 1$, where the caret denotes $\overline{\rm MS}$ subtraction and $\hat m_q \equiv \hat m_q(\hat m_q)$ is the $\overline{\rm MS}$ mass. The zeroth moment is defined separately, through the opposite limit $q^2\to-\infty$ of the dispersion relation rather than as a derivative at $q^2=0$, and requires an explicit ultraviolet subtraction of the asymptotic behavior of $\Pi(q^2)$ \cite{Erler:2002bu,Erler:2016atg,Erler:2022mzd},
\begin{align}
\mathcal{M}_0 :&= -\lim_{t\to\infty}\left[\hat\Pi_q(-t) - \hat\Pi_q^\infty(-t)\right] \nonumber \\
&= \int_{\hat m_q^2}^{\infty} \frac{ds}{s}\left[R_q(s) - R_q^\infty(s)\right].
\label{eq:M0_def}
\end{align}
Because it originates from the ultraviolet limit rather than the low-energy expansion, the zeroth moment carries an enhanced sensitivity to the continuum region above the narrow resonances, a feature that will be exploited throughout this work.

\subsection{Hadronic Side}

The hadronic spectral function is decomposed as
\begin{equation}
R_q(s) = R_q^{\rm res}(s) + R_q^{\rm cont}(s).
\label{eq:Rq_decomp}
\end{equation}
Below the open heavy-quark threshold, the cross section is dominated by a small number of narrow resonances, approximated by $\delta$-functions,
\begin{equation}
R_q^{\rm res}(s) = \sum_R \frac{9\pi\,\Gamma^e_R}{M_R\,\alpha_{\rm em}^2(M_R)}\,\delta(s-M_R^2),
\label{eq:Rres}
\end{equation}
with $M_R$ and $\Gamma^e_R$ the mass and electronic width of resonance $R$ \cite{ParticleDataGroup:2026aaa}, and $\alpha_{\rm em}(M_R)$ the running fine-structure constant at the resonance mass. The resonance parameters used in this analysis are listed in Table~\ref{tab:resonances} and the associated uncertainties, which are dominated by the electronic widths, are propagated to the final result as a statistical uncertainty. Since the electronic widths of $\Upsilon(1S)$, $\Upsilon(2S)$, and $\Upsilon(3S)$ are dominated by a common underlying measurement~\cite{Dehnadi:2011gc}, we conservatively assign half of each width's uncertainty as correlated among these three states, treating the remaining half, and the uncertainties of $\Upsilon(4S)$ and $\Upsilon(5S)$ as uncorrelated. By analogy, since $J/\psi$ and $\psi(2S)$ are likewise dominated by a common experimental program, we adopt the same prescription for the charm sector, treating $\psi(3770)$ and $\psi(4040)$, obtained from independent line-shape analyses, as uncorrelated with the rest.
\begin{table}[h!]
\caption{\label{tab:resonances}%
Experimental input for the charmonium and bottomonium resonances
used in this work~\cite{ParticleDataGroup:2026aaa}: from left to right, resonance name, mass in GeV, electronic decay width in KeV, total decay width in MeV, and ratio of the electromagnetic decay constant considered at $\sqrt{s}=0$ and $\sqrt{s}=M_R$.}
\centering
\bgroup
\def\arraystretch{2.}
\setlength{\tabcolsep}{2pt}
\sisetup{
  separate-uncertainty = false,
}
\begin{tabular}{
  |c|
  S[table-format=2.6(2)]|
  S[table-format=1.3(2)]|
  S[table-format=2.4(3)]|
  S[table-format=1.6]|
}
  \hline
  {Res.} & {$M_R\;[\mathrm{GeV}]$} & {$\Gamma_R^e\;[\mathrm{keV}]$}
  & {$\Gamma_R^{\mathrm{tot}}\;[\mathrm{MeV}]$}
  & {$\dfrac{\alpha_{\mathrm{em}}^2(0)}{\alpha_{\mathrm{em}}^2(M_R)}$} \\
  \hline
  \multicolumn{5}{|c|}{\textit{Charmonium}} \\
  \hline
  $J/\psi(1S)$ & 3.0969  & 5.53(10)   & 0.0926(17) & 0.9532 \\
  $\psi(2S)$   & 3.6861 & 2.33(4)    & 0.286(16)  & 0.9526 \\
  $\psi(3770)$ & 3.7737(7)    & 0.256(16)  & 27.5(9)    & 0.9520 \\
  $\psi(4040)$ & 4.0396(43)   & 0.86(7)    & 84.5  (12.3)  & 0.9500 \\
  \hline
  \multicolumn{5}{|c|}{\textit{Bottomonium}} \\
  \hline
  $\Upsilon(1S)$ & 9.4604(1)  & 1.340(18) & 0.0540(13) & 0.9313 \\
  $\Upsilon(2S)$ & 10.0234(5)   & 0.612(11) & 0.0320(26) & 0.9301 \\
  $\Upsilon(3S)$ & 10.3551(5)   & 0.443(8)  & 0.0203(19) & 0.9295 \\
  $\Upsilon(4S)$ & 10.5794(12)  & 0.272(29) & 20.5(25)     & 0.9290 \\
  $\Upsilon(5S)$ & 10.8852(26)  & 0.310(70) & 37.0(40)     & 0.9284 \\
  \hline
\end{tabular}
\egroup
\end{table}

Above threshold, the continuum is described by a parametric ansatz constructed to interpolate smoothly between the open heavy-quark pair-production threshold and the massless pQCD prediction, which it reproduces asymptotically. Relative to the ansatz used in Refs.~\cite{Erler:2016atg,Erler:2022mzd}, we retain mass corrections to the continuum up to and including $O(\hat m_q^4/s^2)$ \cite{Dehnadi:2011gc,Chetyrkin:2000zk}, together with their associated $O(\alpha_s^3)$ coefficients,
\begin{widetext}
\begin{align}
&R_q^{\mathrm{cont}}(s)
  = 3Q_q^2\,\lambda_1^q(s)
    \sqrt{1-\frac{4\hat{m}_q^2(2M)}{s'}}
    \Bigg(1 + \lambda_3^q \Bigg[ \frac{2\hat{m}_q^2(\sqrt{s})}{s'}
\nonumber\\
&+ \frac{2\hat{m}_q^2(\sqrt{s})}{s'}\Bigg[
  6\,\frac{\alpha_s}{\pi}
  + \left(57.250 - 2.167\,n_f\right)\left(\frac{\alpha_s}{\pi}\right)^{\!2}
  + \left(446.906 - 49.225\,n_f + 0.609\,n_f^2\right)\left(\frac{\alpha_s}{\pi}\right)^{\!3}
\Bigg]
\nonumber\\
&+ \left(\frac{2\hat{m}_q^2(\sqrt{s})}{s'}\right)^2 \Bigg[
  2\,\frac{\alpha_s}{\pi}
  + \left(101.283 -2.800\, n_f + \left( -1.625+0.083\, n_f \right) L_m\right)\left(\frac{\alpha_s}{\pi}\right)^{\!2}
  \nonumber\\
&\qquad\qquad
  + \Big(1595.098 - 119.828\, n_f + 1.845\, n_f^2 + \left( -59.528+4.463\, n_f -0.046\, n_f^2 \right) L_m
  \nonumber\\
&\qquad\qquad\quad
  +  \left( 3.25 -0.167 \, n_f \right) L_m^2
  \Big)\left(\frac{\alpha_s}{\pi}\right)^{\!3}
\Bigg]\Bigg]\Bigg),
\label{eq:Rcont}
\end{align}
\end{widetext}
with $s' = s + 4(\hat m_q^2 - M^2)$, $M$ the mass of the lightest pseudoscalar heavy-flavor meson, $L_m \equiv \log(\hat m_q^2/s)$, and $\lambda_3^q$ a free shape parameter fixed by the sum-rule self-consistency condition. $\lambda_1^q(s)$ is the massless perturbative-QCD correction factor, known to $O(\alpha_s^3)$. Writing $\hat a_s=\hat\alpha_s(\sqrt{s})/\pi$, the $\overline{\rm MS}$ strong-coupling constant, one has
\begin{eqnarray}
  \lambda_1^q(s)
  &=& 1+\hat a_s + \frac{3Q_q^2 \alpha_{\mathrm{em}}}{4\pi} \left( 1 - \frac{1}{3} \hat a_s \right)
  \nonumber\\
  &&+\hat a_s^2\left[
    \frac{365}{24}-11\zeta(3)
    +n_q\left(\frac{2}{3}\zeta(3)-\frac{11}{12}\right)
  \right]
  \nonumber\\
  &&+\hat a_s^3\bigg[
    \frac{87029}{288}-\frac{121}{8}\zeta(2)
    -\frac{1103}{4}\zeta(3)
    +\frac{275}{6}\zeta(5)
  \nonumber\\
  &&\hspace{0.35cm}
    +n_q\left(
      -\frac{7847}{216}+\frac{11}{6}\zeta(2)
      +\frac{262}{9}\zeta(3)-\frac{25}{9}\zeta(5)
    \right)
  \nonumber\\
  &&\hspace{0.35cm}
    +n_q^2\left(
      \frac{151}{162}-\frac{1}{18}\zeta(2)
      -\frac{19}{27}\zeta(3)
    \right)
  \bigg],
  \label{eq:lambda-one}
\end{eqnarray}
\noindent
where $n_q=n_l+1$ and $n_l$ is the number of light flavors. As discussed below, the value of $\lambda_3^q$ obtained from the sum rules is also checked against an independent, data-driven determination, $\lambda_3^{q,\rm exp}$. The deviation between the two, together with the experimental uncertainty on $\lambda_3^{q,\rm exp}$, is included as a separate contribution to the error budget, with the full extraction of $\lambda_3^{q,\rm exp}$ deferred to Sec.~\ref{sec:expdata}.

Two aspects of this $\textit{ansatz}$ differ from previous implementations of the method. First, in the bottom-quark analysis of Ref.~\cite{Erler:2022mzd}, explicit modeling of the $\Upsilon(4S)$ and $\Upsilon(5S)$ states on top of the continuum was introduced to control the determination of $\lambda_3^b$, motivated by the non-trivial resonant structure just above threshold; an analogous treatment had not previously been tested for charm. We extend the explicit resonance content of the charm $\textit{ansatz}$ to include the $\psi(3770)$ and $\psi(4040)$ states, rather than absorbing them into the continuum, and adopt four explicit resonances as the default scenario for the charm contribution. The effect of this choice on the charm-mass determination is examined in Sec.~\ref{sec:results} (Fig.~\ref{fig:NRES4}). Second, the mass corrections to the continuum $\textit{ansatz}$ given in Eq.~\eqref{eq:Rcont} extend the treatment of Ref.~\cite{Erler:2022mzd}, which retained only the term linear in $\lambda_3^q\,\hat m_q^2/s$, to $O(\hat m_q^4/s^2)$ together with the associated $O(\alpha_s^3)$ corrections; the impact of this extension on the charm and bottom determinations is likewise discussed in Sec.~\ref{sec:results} (Fig.~\ref{fig:massive_correction_bottom}).

\subsection{Theoretical side}

On the perturbative side, the moments are obtained from the low-energy expansion of the vector-current correlator. Using $z = q^2/4m_q^2$, with $m_q$ the heavy-quark pole mass, and $a_s \equiv \alpha_s(\mu^2)/\pi$,
\begin{eqnarray}
  \Pi(z)
  &=& \Pi^{(0)}(z)
   +C_F\, \Pi^{(1)}(z)\, a_s
   +\Pi^{(2)}(z) \, a_s^2
  \nonumber\\
  &&{}+\Pi^{(3)}(z)\, a_s^3 + \mathcal{O}\left( a_s^4 \right) \, ,
  \label{eq:Pi_expansion}
\end{eqnarray}
with $C_F=4/3$. Expanding around $z=0$ at renormalization scale $\mu=m_q$,
\begin{equation}
\Pi^{(i)}(z) = \frac{3Q_q^2}{16\pi^2}\sum_{n=1}^{\infty} C_n^{(i)}\, z^n,
\label{eq:Pi_lowenergy}
\end{equation}
the theoretical moments in the pole scheme read
\begin{equation}
\mathcal{M}_n^{\rm pQCD} = \frac{9Q_q^2}{4}\left(\frac{1}{2m_q}\right)^{2n} C_n,
\label{eq:Mn_pQCD}
\end{equation}
where
\begin{equation}
C_n = C_n^{(0)} + C_F\,C_n^{(1)}\,a_s + C_n^{(2)}\,a_s^2 + C_n^{(3)}\,a_s^3 + O(a_s^4).
\label{eq:Cn}
\end{equation}
The quark-mass dependence factors out of the integral entirely, so the coefficients $C_n^{(i)}$ \cite{Chetyrkin:1997mb,Maier:2007yn,Kiyo:2009gb,Greynat:2011zp}, collected in Table~\ref{tab:Cnk}, need be computed only once; where these coefficients are not known exactly, their quoted uncertainties are also independently propagated to the final error budget. The relation between the pole mass $m_q$ and the $\overline{\rm MS}$ mass $\hat m_q$ follows from the known on-shell$\,$--$\,\overline{\rm MS}$ matching relations. Up to ${\cal O}(\alpha_s^3)$, one finds~\cite{Tarrach:1980up,Fleischer:1998dw,Broadhurst:1993mw,Ball:1995ni,Chetyrkin:1999qi,Melnikov:2000qh}
\begin{eqnarray}
  m_q &=& \hat m_q(\hat m_q)\left[1+1.333\,\hat a_s
  +\hat a_s^2\left(-1.041\,n_l+13.443\right)\right.
  \nonumber\\
  &&\left.
  +\hat a_s^3\left(0.653\,n_l^2-26.655\,n_l+190.595\right)
  \right],
  \label{eq:mass-rge}
\end{eqnarray}
with
\begin{eqnarray}
  a_s\equiv\frac{\alpha_s(m_q)}{\pi}
  =\hat a_s+\frac{4n_l-62}{9}\,\hat a_s^3.
  \label{eq:alpha-rge}
\end{eqnarray}

The leading non-perturbative correction, from the dimension-four gluon condensate $\langle\alpha_s G^2/\pi\rangle$, contributes to the moments as \cite{Novikov:1977dq}
\begin{equation}
\mathcal M_n^{\rm cond} = \frac{12\pi^2 Q_q^2}{(4\hat m_q^2)^{n+2}}\left\langle\frac{\alpha_s}{\pi}G^2\right\rangle a_n\left(1+\frac{\alpha_s(\hat m_q)}{\pi}\hat b_n\right),
\label{eq:Mcond}
\end{equation}
with
\begin{eqnarray}
  a_n &=& -\frac{2n+2}{15} \frac{\Gamma(4+n)\,\Gamma\!\left(\frac{7}{2}\right)}{\Gamma(4)\,\Gamma\!\left(\frac{7}{2}+n\right)}\, , \nonumber \\
  \hat{b}_n &=& b_n -\frac{4}{3} (2n+4),
\end{eqnarray}
\noindent
given in \cite{Broadhurst:1994qj,Kuhn:2007vp,Chetyrkin:2010ic}. We adopt $\langle\alpha_s G^2/\pi\rangle = 0.005\,{\rm GeV}^4$ with a conservative $100\%$ uncertainty, following Ref.~\cite{Dominguez:2014fua}; this uncertainty, together with that on $\hat\alpha_s(M_Z)$, is likewise propagated as statistical uncertainty.

Since the perturbative series for the moments is known only to $O(\alpha_s^3)$, the effect of missing higher orders is estimated following Refs.~\cite{Erler:2002bu,Erler:2016atg,Erler:2022mzd}, by retaining the largest group-theoretical factor at the next uncalculated order,
\begin{equation}
\Delta \mathcal{M}_n^{(i)} = \pm Q_q^2\, N_C C_F C_A^{i-1}\left(\frac{\alpha_s(\hat m_q)}{\pi}\right)^{\!i}\left(\frac{1}{2\hat m_q}\right)^{\!2n},
\label{eq:trunc}
\end{equation}
with $N_C=C_A=3$, $C_F=4/3$. This prescription is conservative: applied to the last known order, it systematically overestimates the actual size of that term. Combined with the sources listed above, this truncation uncertainty completes the error budget entering the determination of $\hat m_q$ and $\lambda_3^q$ from a given pair of moments, discussed further in Sec.~\ref{sec:results}.

\section{Generalized Correlated Sum Rules}
\label{sec:generalized}

The theory-side prediction for the moment $\mathcal{M}_n$ is
\begin{equation}
\mathcal{M}_n^{\rm th} \equiv \mathcal{M}_n^{\rm pQCD} + \mathcal{M}_n^{\rm cond},
\label{eq:Mn_th}
\end{equation}
and the hadronic-side prediction is
\begin{equation}
\mathcal{M}_n^{\rm had} \equiv \int_{4\hat m_q^2}^{\infty} \frac{ds}{s^{n+1}}\left[R_q^{\rm res}(s) + R_q^{\rm cont}(s)\right].
\label{eq:Mn_had}
\end{equation}
Imposing these to agree,
\begin{equation}
\mathcal{M}_n^{\rm th} \overset{!}{=} \mathcal{M}_n^{\rm had},
\label{eq:consistency}
\end{equation}
for two different values of $n$ determines $\hat m_q$ and $\lambda_3^q$; the remaining moments then serve as consistency checks.

Following Refs.~\cite{Erler:2002bu,Erler:2016atg,Erler:2022mzd}, the zeroth moment must be one of the two moments used. Unlike the higher moments, which arise as derivatives of the correlator at $Q^2=0$ and are therefore governed mainly by the region near threshold, the zeroth moment originates from the opposite, ultraviolet limit of the dispersion relation, and is correspondingly most sensitive to the continuum. Including it is what breaks the near-degeneracy between $\hat m_q$ and $\lambda_3^q$ that persists if only higher moments are used. We stress that this matching between the hadronic and perturbative descriptions is required only on average, not locally; this is the sense in which quark-hadron duality enters the sum rule framework, a substantially weaker and better justified assumption than the local matching discussed in Sec.~\ref{sec:hvp}.

The quantity $\hat m_q$, extracted as above from equating a pair of sum rules from Eq.~(\ref{eq:consistency}), and the HVP contribution $a_\mu^{(i)}$ from Eq.~(\ref{eq:amu_hvp}) are not independent. Eq.~\eqref{eq:amu_hvp} is itself an integral over the same spectral function $R_q(s)$ that enters the moment sum rules, differing only in its integration kernel. The ordinary moments weight $R_q(s)$ by $s^{-(n+1)}$, while $a_\mu^{(i)}$ weights it by $\hat K^{(i)}(s)/s^2$. Since $\hat m_q$ and $a_\mu^{(i)}$ are both determined by integrals of the same function, any shift in one is necessarily correlated with a shift in the other. Rather than treating this correlation as a nuisance to be propagated after the result, we build it directly into the sum-rule framework itself, by generalizing the moment sum rules of Sec.~\ref{sec:sumrules} to an arbitrary admissible kernel $K(s)$. We define the family of generalized moments
\begin{equation}
a_n[K] := \int_{4\hat m_q^2}^{\infty} \frac{ds}{s^{n+1}}\,K(s)\,R_q(s),
\label{eq:An_def}
\end{equation}
for $n \geq 1$, which for $K(s)=\hat K^{(i)}(s)$ reproduces at $n=1$ exactly the integral in Eq.~\eqref{eq:amu_hvp}, up to the known kinematic prefactor. Just as the ordinary moments arise as Taylor coefficients of the correlator $\Pi(Q^2)$ about $Q^2=0$, the generalized moments can be thought as the expansion coefficients of an associated function
\begin{equation}
\hat{A}(Q^2;K) \equiv A(Q^2;K) - A(0;K) = \sum_{n=1}^{\infty} a_n[K]\,(Q^2)^n,
\label{eq:Ahat_def}
\end{equation}
so that
\begin{equation}
\mathcal{A}_n[K] := \frac{1}{n!}\frac{\partial^n A(Q^2;K)}{\partial (Q^2)^n}\bigg|_{Q^2=0} = \int_{4\hat m_q^2}^{\infty} \frac{ds}{s^{n+1}}\,K(s)\,R_q(s).
\label{eq:An_taylor}
\end{equation}
Setting $K(s)\to 1$ recovers the ordinary moments of Sec.~\ref{sec:sumrules} exactly, so the generalized construction is the same object, with the flat weight replaced by the physical kernel of interest. In what follows $K$ denotes a generic admissible kernel; explicit results are obtained in Sec.~\ref{sec:results} by specializing to $K=\hat K^{(2)},\hat K^{(4a)},\hat K^{(4b)}$.

For $K=1$, the moment $\mathcal{M}_n$ is obtained directly as a low-energy Taylor coefficient of $\Pi(Q^2)$, without even explicitly integrating $\mathrm{Im}\,\Pi(s)$. This is precisely what defines a sum rule, a finite perturbative quantity equated to a hadronic integral, with no integration required on the theoretical side. For a general kernel $K(s)$, this shortcut is not directly available, and in principle the complete integral over the imaginary part of the correlator would as well have to be carried out on the theoretical side. The kernels of interest here, however, are particularly convenient, because $m_\mu^2/(4\hat m_q^2)$ is small, so the high-energy expansion of the HVP kernels is well justified. For charm, $m_\mu^2/(4\hat m_c^2)\approx 10^{-3}$; for bottom, $m_\mu^2/(4\hat m_b^2)\approx 10^{-4}$, making the expansion essentially exact~\cite{Krause:1996rf},
\begin{equation}
\hat K\!\left(\frac{4\hat m_q^2}{m_\mu^2}z\right) \simeq \sum_{j\geq 0}\sum_{r=0}^{2} K_r^{(j)}\,\log^r\!\left(\frac{m_\mu^2}{4\hat m_q^2 z}\right)\left(\frac{m_\mu^2}{4\hat m_q^2 z}\right)^{\!j}.
\label{eq:kernel_expansion}
\end{equation}
whose explicit form is given in Appendix~\ref{app:kernel_expansion}. In practice it suffices to sum up to $j=3$. The identity
\begin{equation}
  \log^r\!\left(\frac{m_\mu^2}{4\hat m_q^2 z}\right)
  =\sum_{k=0}^{r}\binom{r}{k}
    \log^k\!\left(\frac{1}{z}\right)
    \log^{r-k}\!\left(\frac{m_\mu^2}{4\hat m_q^2}\right)\, ,
  \label{eq:logexpand}
\end{equation}
separates the mass dependence from the integration variable as for the ordinary moments, and the theoretical side reduces to a fixed, finite combination of logarithmic moments,
\begin{equation}
C_{n,r}^{(i)} := \int_1^{\infty} \frac{dz}{z^{n+1}}\log^r(z)\,R_q^{(i)}(z).
\label{eq:Cnr_def}
\end{equation}
\noindent
The $r=0$ coefficients are simply the ordinary moments already required for the standard sum rules of Sec.~\ref{sec:sumrules}; only the $r=1,2$ pieces are genuinely new, and their evaluation requires reconstructing the perturbative spectral function order by order in $\alpha_s$, matching its known behavior at the low-energy, threshold, and high-energy limits \cite{Greynat:2010kx,Greynat:2011zp}. They can all be found in Table~\ref{tab:Cnk}. This reorganization is an improvement in precision rather than a complication, and does not compromise the finiteness that defines a sum rule, since the higher-order logarithmic coefficients turn out to be relatively suppressed.

Collecting these pieces, the generalized moment on the pQCD side reads
\begin{eqnarray}
  \mathcal{A}_n^{\rm pQCD}[K]
  &=& \frac{9Q_q^2}{4}
    \left(\frac{1}{2\hat m_q}\right)^{2n}
    \sum_{j\geq0}
    \left(\frac{m_\mu^2}{4\hat m_q^2}\right)^j
  \nonumber\\
  &&\times\bigg[
    \hat D_n^{(0,j)}
    +\hat D_n^{(1,j)}
      \log\!\left(\frac{m_\mu^2}{4\hat m_q^2}\right)
  \nonumber\\
  &&\hspace{0.55cm}
    +\hat D_n^{(2,j)}
      \log^2\!\left(\frac{m_\mu^2}{4\hat m_q^2}\right)
  \bigg],
  \label{eq:An_pQCD}
\end{eqnarray}
where, in the pole scheme,
\begin{equation}
D_n^{(r,j)} = \sum_{p=0}^{2-r} (-1)^p \binom{r+p}{p} K_{r+p}^{(j)}\, C_{n+j,p}.
\label{eq:Dnrj}
\end{equation}
Since the HVP kernels are monotonically increasing functions bounded above by their asymptotic value, $K(s)\leq1$, the ordinary condensate contribution $\mathcal{M}_n^{\rm cond}$, Eq.~\eqref{eq:Mcond}, evaluated at $K=1$, is always an upper bound on the true kernel-weighted contribution. We conservatively retain $\mathcal{A}_n^{\rm cond}\equiv\mathcal{M}_n^{\rm cond}$ unmodified in the generalized sum rule, rather than deriving its kernel-weighted counterpart. The full theory-side prediction entering the generalized sum rule is then
\begin{equation}
\mathcal{A}_n^{\rm th}[K] \equiv \mathcal{A}_n^{\rm pQCD}[K] + \mathcal{A}_n^{\rm cond},
\label{eq:An_th}
\end{equation}
reducing to $\mathcal{M}_n^{\rm pQCD}+\mathcal{M}_n^{\rm cond}$, in Eq.~\eqref{eq:Mn_th}, for $K\to1$. That is, setting $K\to 1$ collapses the sum in Eq.~\eqref{eq:An_pQCD} to its $j=0$ term and reproduces Eq.~\eqref{eq:Mn_pQCD} exactly.

The generalized sum rule is the equality between this perturbative expansion and the corresponding hadronic integral,
\begin{equation}
\mathcal{A}_n^{\rm had}[K] \equiv \int_{4\hat m_q^2}^{\infty} \frac{ds}{s^{n+1}}\,K(s)\left[R_q^{\rm res}(s) + R_q^{\rm cont}(s)\right],
\label{eq:An_had}
\end{equation}
imposed, exactly as in Sec.~\ref{sec:sumrules},
\begin{equation}
    \mathcal{A}_n^{\rm th}[K] \overset{!}{=} \mathcal{A}_n^{\rm had}[K],
\label{eq:generalized_sumrule}
\end{equation}
for a pair of moments including the zeroth moment. Taking $K=\hat K^{(i)}$ and $n=1$ in place of an ordinary higher moment, this condition is no longer an auxiliary relation from which $a_\mu^{(i)}$ must subsequently be derived. It already is, at that order, the direct statement of consistency between the theory-side and hadronic-side descriptions of $a_\mu^{(i)}$ itself. Using a pair including the $n=1$ generalized moment, $\mathcal{A}_1[\hat K^{(i)}]$, together with the zeroth moment, to fix $\hat m_q$ and $\lambda_3^q$ makes the two descriptions of $a_\mu^{(i)}$ coincide exactly by construction. The remaining generalized moments, at $n\neq 1$ or evaluated instead with $K=1$, are not automatically matched, and the residual spread between their theory-side and hadronic-side evaluations provides a direct, physically meaningful measure of the duality-violation and model systematics specific to $a_\mu^{(i)}$, as shown in Sec.~\ref{sec:results}.

The construction of the zeroth moment itself must be revisited when $K\neq 1$. The HVP kernels $\hat K^{(i)}$ are analytic in the complex $s$-plane with a cut along the negative real axis, and satisfy the dispersion relation~\cite{Barbieri:1974nc,Balzani:2021del}
\begin{equation}
\hat K(s/m_\mu^2) = 1 + \frac{1}{\pi}\int_{-\infty}^{0} ds'\,\frac{{\rm Im}\,\hat K(s'/m_\mu^2)}{s'-s}\, ,
\label{eq:kernel_disp}
\end{equation}
for $s>0$. Using this relation to rewrite the $K$-weighted zeroth integral, and exchanging the order of integration, the second term can be expressed entirely as a spacelike integral of the correlator, evaluated away from the physical cut,
\begin{equation}
\mathcal{A}_0^{\rm pQCD}[\hat K] = \mathcal{M}_0^{\rm pQCD} - 12\pi\int_{-\infty}^{0} \frac{dz}{z}\,{\rm Im}\,\hat K\!\left(\frac{4\hat m_q^2}{m_\mu^2}z\right)\hat\Pi(z),
\label{eq:A0_generalized}
\end{equation}
where $\mathcal{M}_0^{\rm pQCD}$ is the standard zeroth moment defined in Eq.~\eqref{eq:M0_def} and the additional piece is a controlled Euclidean integral, evaluated in pQCD at the same order in $\alpha_s$ as the rest of the analysis. This second term in Eq. \eqref{eq:A0_generalized} is numerically suppressed relative to $\mathcal{M}_0^{\rm pQCD}$, and it is therefore not considered further for our present purposes, which scope at the per-mil level.

\section{Results}
\label{sec:results}

\subsection{Constrained sum rules results}\label{SubSec:ConstrainedSR}
\begin{table*}[t]
\caption{\label{tab:errorbudget_charm}
Charm-sector results for $\hat m_c$ and the leading-order contribution to
$a_\mu^c({\rm LO})$ (in units of $10^{-10}$), obtained from the sum rule weighted by $\hat K^{(2)}$, for the moment pairs $(\mathcal{M}_0,\mathcal{M}_1)$, $(\mathcal{M}_0,\mathcal{M}_2)$, and $(\mathcal{M}_0,\mathcal{M}_3)$, together with the breakdown of the uncertainty sources discussed in Sec.~\ref{sec:sumrules}. $a_\mu^{c,{\rm th}}$ denotes the theory-side determination using Eq.\eqref{eq:An_th}  and $a_\mu^{c,{\rm had}}$ the hadronic-side determination using Eq. \eqref{eq:An_had}. The adopted default, $(\mathcal{M}_0,\mathcal{M}_2)$, is shown in bold.}
\centering
\renewcommand{\arraystretch}{1.3}
\setlength{\tabcolsep}{4pt}
\sisetup{
  separate-uncertainty = false,
  table-align-uncertainty = false,
}
\begin{tabular}{|l|
  S[table-format=4.1]|
  S[table-format=2.4]|
  S[table-format=2.4]|
  >{\bfseries}S[table-format=4.1]|
  >{\bfseries}S[table-format=2.4]|
  >{\bfseries}S[table-format=2.4]|
  S[table-format=4.1]|
  S[table-format=2.4]|
  S[table-format=2.4]|
}
\hline
& \multicolumn{3}{c|}{$0^{\rm th}+1^{\rm st}$} & \multicolumn{3}{c|}{$\mathbf{0^{\rm th}+2^{\rm nd}}$} & \multicolumn{3}{c|}{$0^{\rm th}+3^{\rm rd}$} \\
\hline
{Source} & {$\hat m_c$~[\mdseries MeV]} & {$a_\mu^{c,{\rm th}}$} & {$a_\mu^{c,{\rm had}}$}
         & {$\hat m_c$~[\mdseries MeV]} & {$a_\mu^{c,{\rm th}}$} & {$a_\mu^{c,{\rm had}}$}
         & {$\hat m_c$~[\mdseries MeV]} & {$a_\mu^{c,{\rm th}}$} & {$a_\mu^{c,{\rm had}}$} \\
\hline
Central value & {1264.1} & {14.5111} & {14.5111} & {1267.1} & {14.4382} & {14.4896} & {1268.1} & {14.4137} & {14.4823} \\
\hline
Truncation & {4.6} & {0.0030} & {0.0030} & {5.2} & {0.0201} & {0.0080} & {6.3} & {0.0464} & {0.0158} \\
$\Delta C_{n,r}^{(i)}$ & {0.05} & {0.0004} & {0.0004} & {0.002} & {0.0008} & {0.0000} & {0.04} & {0.0000} & {0.0003} \\
$\lambda_3^q \neq \lambda_3^{q,\rm exp}$ & {1.2} & {0.0298} & {0.0298} & {0.6} & {0.0153} & {0.0402} & {0.3} & {0.0072} & {0.0437} \\
$\Delta\lambda_3^{q,\rm exp}$ & {4.3} & {0.1060} & {0.1060} & {1.6} & {0.0397} & {0.1041} & {0.7} & {0.0172} & {0.1035} \\
Statistical & {4.8} & {0.0957} & {0.0957} & {3.9} & {0.0945} & {0.0940} & {3.7} & {0.1059} & {0.0941} \\
\hline
Total & {8.0} & {0.1459} & {0.1459} & \bfseries{6.8} & \bfseries{0.1056} & \bfseries{0.1461} & {7.4} & {0.1172} & {0.1474} \\
\hline
\end{tabular}
\end{table*}

\begin{table*}[t]
\caption{\label{tab:errorbudget_bottom}
Bottom-sector results for $\hat m_b$ and the leading-order contribution to
$a_\mu^b({\rm LO})$ (in units of $10^{-10}$), obtained from the sum rule weighted by $\hat K^{(2)}$, for the moment pairs $(\mathcal{M}_0,\mathcal{M}_5)$, $(\mathcal{M}_0,\mathcal{M}_6)$, and $(\mathcal{M}_0,\mathcal{M}_7)$, together with the breakdown of the uncertainty sources discussed in Sec.~\ref{sec:sumrules}. $a_\mu^{b,{\rm th}}$  denotes the theory-side determination using Eq.\eqref{eq:An_th}  and $a_\mu^{b,{\rm had}}$ the hadronic-side determination using Eq. \eqref{eq:An_had}. The adopted default, $(\mathcal{M}_0,\mathcal{M}_6)$, is shown in bold.}
\centering
\renewcommand{\arraystretch}{1.3}
\setlength{\tabcolsep}{4pt}
\sisetup{
  separate-uncertainty = false,
  table-align-uncertainty = false,
}
\begin{tabular}{|l|
  S[table-format=4.1]|
  S[table-format=1.5]|
  S[table-format=1.5]|
  >{\bfseries}S[table-format=4.1]|
  >{\bfseries}S[table-format=1.5]|
  >{\bfseries}S[table-format=1.5]|
  S[table-format=4.1]|
  S[table-format=1.5]|
  S[table-format=1.5]|
}
\hline
& \multicolumn{3}{c|}{$0^{\rm th}+5^{\rm th}$} & \multicolumn{3}{c|}{$\mathbf{0^{\rm th}+6^{\rm th}}$} & \multicolumn{3}{c|}{$0^{\rm th}+7^{\rm th}$} \\
\hline
{Source} & {$\hat m_b$~[\mdseries MeV]} & {$a_\mu^{b,{\rm th}}$} & {$a_\mu^{b,{\rm had}}$}
         & {$\hat m_b$~[\mdseries MeV]} & {$a_\mu^{b,{\rm th}}$} & {$a_\mu^{b,{\rm had}}$}
         & {$\hat m_b$~[\mdseries MeV]} & {$a_\mu^{b,{\rm th}}$} & {$a_\mu^{b,{\rm had}}$} \\
\hline
Central value & {4181.8} & {0.30132} & {0.30066} & {4182.3} & {0.30126} & {0.30063} & {4182.7} & {0.30120} & {0.30061} \\
\hline
Truncation & {3.8} & {0.00028} & {0.00013} & {4.4} & {0.00038} & {0.00017} & {5.2} & {0.00050} & {0.00021} \\
$\Delta C_{n,r}^{(i)}$ & {1.3} & {0.00022} & {0.00009} & {2.4} & {0.00037} & {0.00014} & {3.5} & {0.00055} & {0.00021} \\
$\lambda_3^q \neq \lambda_3^{q,\rm exp}$ & {4.6} & {0.00069} & {0.00004} & {2.8} & {0.00044} & {0.00003} & {1.8} & {0.00029} & {0.00002} \\
$\Delta\lambda_3^{q,\rm exp}$ & {1.0} & {0.00015} & {0.00167} & {0.6} & {0.00009} & {0.00167} & {0.4} & {0.00006} & {0.00167} \\
Statistical & {4.9} & {0.00108} & {0.00146} & {4.3} & {0.00106} & {0.00143} & {3.9} & {0.00107} & {0.00141} \\
\hline
Total & {7.8} & {0.00134} & {0.00223} & \bfseries{7.2} & \bfseries{0.00127} & \bfseries{0.00221} & {7.6} & {0.00133} & {0.00220} \\
\hline
\end{tabular}
\end{table*}

\begin{figure}[t]
\centering
\includegraphics[width=\columnwidth]{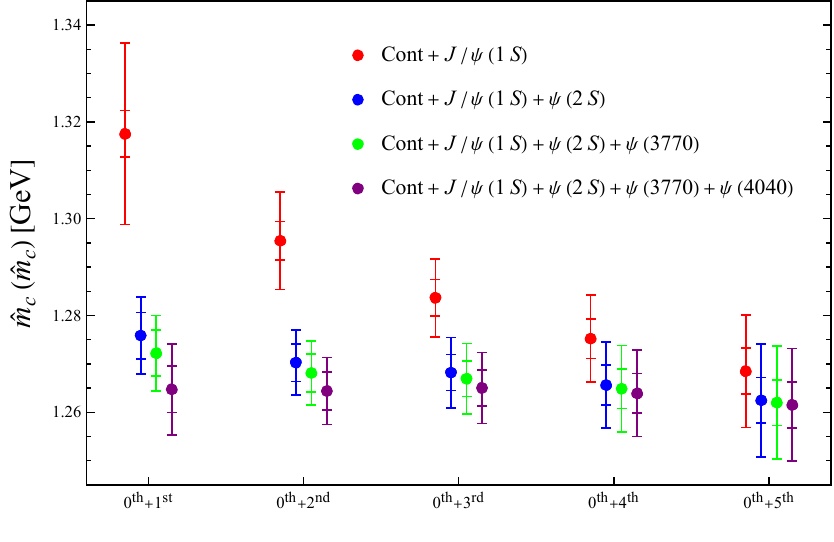}
\caption{\label{fig:NRES4}
Charm-quark mass $\hat m_c(\hat m_c)$ obtained from different combinations of moments $(0^{\rm th}{+}n^{\rm th})$, for an increasing number of explicit narrow resonances included in $R_c^{\rm res}(s)$, Eq. \eqref{eq:Rres}.}
\end{figure}

\begin{figure}[t]
\centering
\includegraphics[width=\columnwidth]{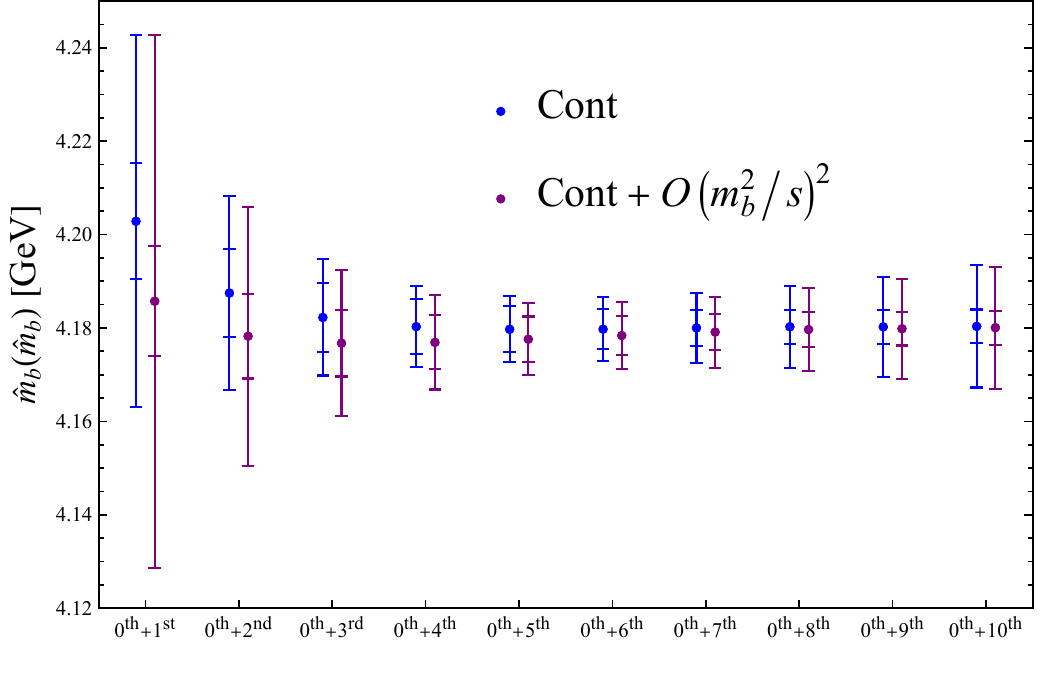}
\caption{\label{fig:massive_correction_bottom}
Bottom-quark mass $\hat m_b(\hat m_b)$ obtained from different combinations of moments $(0^{\rm th}{+}n^{\rm th})$, comparing the continuum ansatz including mass corrections up to $O(\hat m_b^2/s)$ with the extended treatment up to $O(\hat m_b^4/s^2)$ including the associated $O(\alpha_s^3)$ corrections.}
\end{figure}

Before presenting the final determinations for the heavy-quark contributions to the $a_{\mu}^{\rm HVP}$, we comment on the two extensions of the continuum ansatz introduced in Sec.~\ref{sec:sumrules} which go beyond the method used in Refs.~\cite{Erler:2002bu,Erler:2016atg,Erler:2022mzd}: 
\begin{itemize}
    \item Figure~\ref{fig:NRES4} shows $\hat m_c(\hat m_c)$ as a function of the moment pair used, for an increasing number of explicit narrow resonances in $R_c^{\rm cont}(s)$ (or equivalently in $R_c^{\rm res}(s)$). With one or two resonances treated explicitly, the extracted value of $\hat m_c$ still depends visibly on the moment pair; this dependence largely disappears once the $\psi(3770)$ and $\psi(4040)$ are included, and the central value becomes essentially stable across moment pairs. We therefore adopt four explicit resonances as the default for charm throughout this work.
    \item Figure~\ref{fig:massive_correction_bottom} shows the analogous comparison for $\hat m_b(\hat m_b)$ between the continuum ansatz truncated at $O(\hat m_b^2/s)$ and the extended treatment including corrections up to and including $O(\hat m_b^4/s^2)$. The effect is negligible for charm, but for bottom the extended model noticeably improves the stability of $\hat m_b$ across moment pairs, and is adopted as the default (also for charm) in what follows. The fourth and fifth resonances are considered in $R_b^{\rm cont}(s)$ following~\cite{Erler:2022mzd}.
\end{itemize}

The determinations of $\hat m_q$ and $a_\mu^{(i)}$ obtained in this work are not independent quantities. Both follow from the same self-consistency condition, Eq.~\eqref{eq:generalized_sumrule}, applied to the same pair of moments, so a given moment pair fixes $\hat m_q$ and $a_\mu^{(i)}$ simultaneously and with a definite correlation between them. Therefore, a value for $a_\mu^{(i)}$ is only meaningful together with its corresponding value of $\hat m_q$, and the two cannot be varied independently. This correlation is not a byproduct of the analysis but the central feature exploited throughout, and it is what allows the anticorrelation between $\hat m_q$ and $a_\mu^{(i)}$ to reduce the final uncertainty in the latter. We show first such correlation effects for the charm in detail, for latter summarize the bottom counterpart in brief.

\begin{figure}[t]
\centering
\begin{subfigure}[b]{\columnwidth}
\centering
\includegraphics[width=\columnwidth]{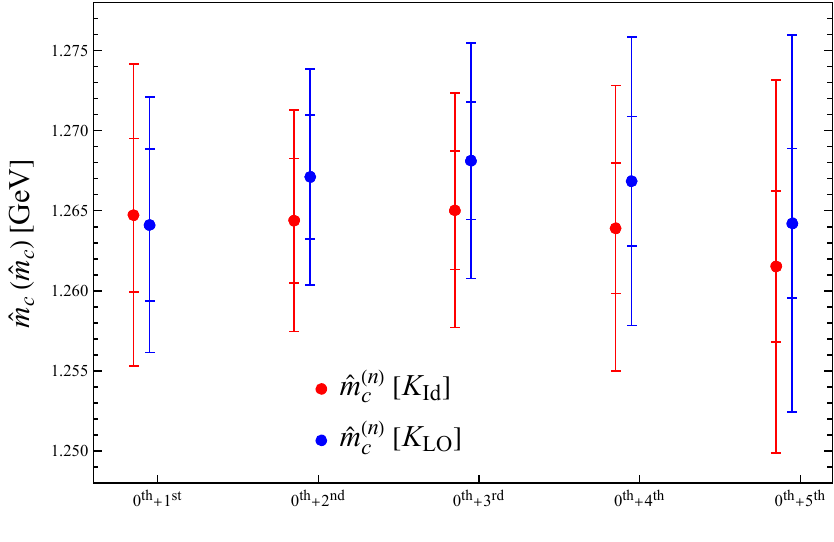}
\caption{$\hat m_c(\hat m_c)$ using different moment pairs and kernels.}
\label{fig:mc}
\end{subfigure}
\\[0.5em]
\begin{subfigure}[b]{\columnwidth}
\centering
\includegraphics[width=\columnwidth]{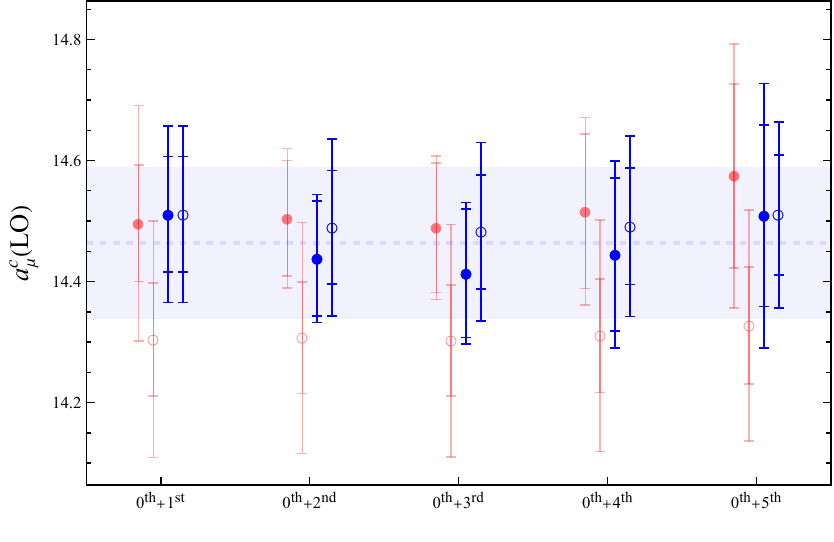}
\caption{$a_\mu^{c}({\rm LO})\times 10^{10}$ across different pairs and kernels.}
\label{fig:amuLOc}
\end{subfigure}
\caption{(a) $\hat m_c(\hat m_c)$ and (b) $a_\mu^{c}({\rm LO})\times 10^{10}$, both as a function of the moment pair $(0^{\rm th}{+}n^{\rm th})$, obtained from the sum-rule self-consistency condition, Eq.~\eqref{eq:generalized_sumrule}, with the identity kernel (red) and with the kernel of the observable, $\hat K^{(2)}$ (blue). Panel (b) additionally distinguishes the theory-side (circles) and hadronic-side (empty points) evaluations of $a_\mu^{c}({\rm LO})$ for the mass shown in panel (a).}
\label{fig:mc_amuLOc}
\end{figure}

Figure~\ref{fig:mc} shows the resulting $\hat m_c$ as a function of the moment pair, for the two kernel choices; figure \ref{fig:amuLOc} shows how this same determination translates into $a_\mu^{c}({\rm LO})$, comparing the two choices of kernel: the identity kernel, $K=1$ (red), and the kernel of the observable itself, $K=\hat K^{(2)}$ (blue). For each kernel and each moment pair $(0^{\rm th}{+}n^{\rm th})$, the self-consistency condition of Eq.~\eqref{eq:generalized_sumrule} is solved for $(\hat m_c,\lambda_3^c)$. Geometrically, each moment equation, together with its uncertainty, defines a band in the $(\hat m_q,\lambda_3^q)$ plane rather than a single curve, and a given pair of moments fixes $(\hat m_q,\lambda_3^q)$ at the intersection of two such bands (c.f., Fig.~\ref{fig:bandplots_charm} in Appendix~\ref{App:charmdata}, for the charm case). Two evaluations of $a_\mu^{c}({\rm LO})$ then follow from the same extracted pair: a theory-side evaluation (circle), obtained by inserting $\hat m_c$ into Eq.~\eqref{eq:An_pQCD}, the generalized analogue of Eq.~\eqref{eq:Mn_pQCD}, combined with the condensate contribution of Eq.~\eqref{eq:Mcond}; and a hadronic-side evaluation (empty point), obtained by inserting $(\hat m_c,\lambda_3^c)$ into the same $\hat K^{(2)}(s)/s^2$-weighted integral of $R_c^{\rm res}(s)+R_c^{\rm cont}(s)$, Eqs.~\eqref{eq:Rres} and~\eqref{eq:Rcont}.

These results demand a careful walkthrough. Consider first the leftmost pair, $0^{\rm th}{+}1^{\rm st}$ in Fig. \ref{fig:amuLOc}. The red solid circle is the theory-side evaluation of $a_\mu^{c}({\rm LO})$ using $\hat m_c$ from the $K=1$ solution of $(\mathcal{M}_0,\mathcal{M}_1)$; the red empty point is the hadronic-side evaluation using the same $\hat m_c,\lambda_3^c$. Because $K=1$ was used to fix the mass, and $\mathcal{M}_1$ in Eq.~\eqref{eq:Mn_def}, carries no direct relation to $a_\mu^{c}({\rm LO})$, i.e. Eq.~\eqref{eq:amu_hvp}, there is no reason for the two evaluations (solid and empty circles) to agree, and indeed a visible offset is already present at this first pair. Even though we use the same kernel to obtain the mass, to obtain $a_\mu^{c}({\rm LO})$ we use another kernel, and thus, potential systematic uncertainties emerge with agreement at the $1\sigma$ level.

Now consider the blue points at the same pair $0^{\rm th}{+}1^{\rm st}$. The blue solid circle is the theory-side evaluation of $a_\mu^{c}({\rm LO})$ using $\hat m_c$ extracted instead from $(\mathcal{A}_0[\hat K^{(2)}],\mathcal{A}_1[\hat K^{(2)}])$, i.e., using the physical kernel already at the calibration step, Eq.~\eqref{eq:An_def}; the blue empty point is the hadronic-side evaluation with the same $(\hat m_c,\lambda_3^c)$. Since $\mathcal{A}_1[\hat K^{(2)}]$ is, up to the known kinematic prefactor of Eq.~\eqref{eq:amu_hvp}, $a_\mu^{c}({\rm LO})$ itself, the moment pair used for calibration is now \textit{literally} the statement that the theory and hadronic descriptions of the observable agree, so the blue solid circle and blue empty point coincide exactly by construction at $0^{\rm th}{+}1^{\rm st}$. This is going to happen only at this pair, the one which mathematically corresponds to the observable.

The same construction is repeated at every other moment pair shown in the figure. Beyond $0^{\rm th}{+}1^{\rm st}$, agreement between the solid circle and the empty point is no longer guaranteed for either kernel, and the residual offset reflects how at most the continuum ansatz,  quark-hadron duality, and unknown perturbative effects depart from the calibration point. This residual offset remains visibly smaller ($\sim 1\sigma$ for the identity kernel, $< \frac12 \sigma$ for $\hat K^{(2)}$), thus compatible with zero within the quoted uncertainties, across the full range of moment pairs shown. The compelling consistency within $1\sigma$ shown in Fig. \ref{fig:amuLOc} is not only a validation of the model down to subtle sub-dominant features but also a precise calibration of the systematic uncertainties thanks to the small observed deviations.

The blue band, obtained at the adopted moment pair as the average of the central values and of the uncertainties of the two evaluations, summarizes the result, and constitutes our preferred determination of the charm-sector contribution to $a_\mu^{c}({\rm LO})$ as it generously covers such aforementioned deviations. The full breakdown of uncertainty sources for the first three moment pairs shown in Fig.~\ref{fig:amuLOc} is given in Table~\ref{tab:errorbudget_charm}. As final result at a given moment pair, we adopt the combination of the theory-side and hadronic-side evaluations of $a_\mu^{q,(i)}$, since this provides the most conservative determination, incorporating not only the sources of uncertainty discussed in Sec.~\ref{sec:sumrules} but also the residual duality-violation and continuum-modeling systematic exposed by their mutual spread.

We stress that the results shown here are always reported jointly: $\hat m_q$ and $a_\mu^{q,(i)}$ obtained from a given moment pair are mutually correlated, and the corresponding correlation matrices for each quark sector are given in Tables~\ref{tab:corrmatrix_charm}, \ref{tab:corrmatrix_bottom} of the Appendix \ref{App:CorrelationMatrices}. NLO contributions are not discussed in this section; their treatment is presented separately in Sec.~\ref{sec:nlo_outlook}.

\begin{figure}[t]
\centering
\begin{subfigure}[b]{\columnwidth}
\centering
\includegraphics[width=\columnwidth]{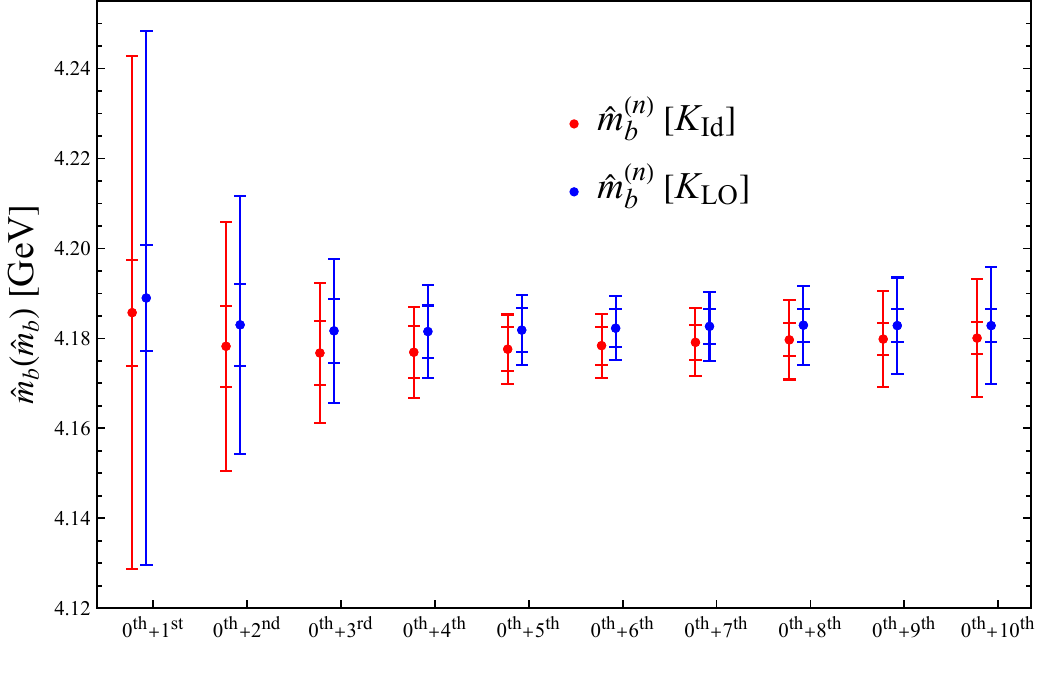}
\caption{$\hat m_b(\hat m_b)$ using different moment pairs and kernels.}
\label{fig:mb}
\end{subfigure}
\\[0.5em]
\begin{subfigure}[b]{\columnwidth}
\centering
\includegraphics[width=\columnwidth]{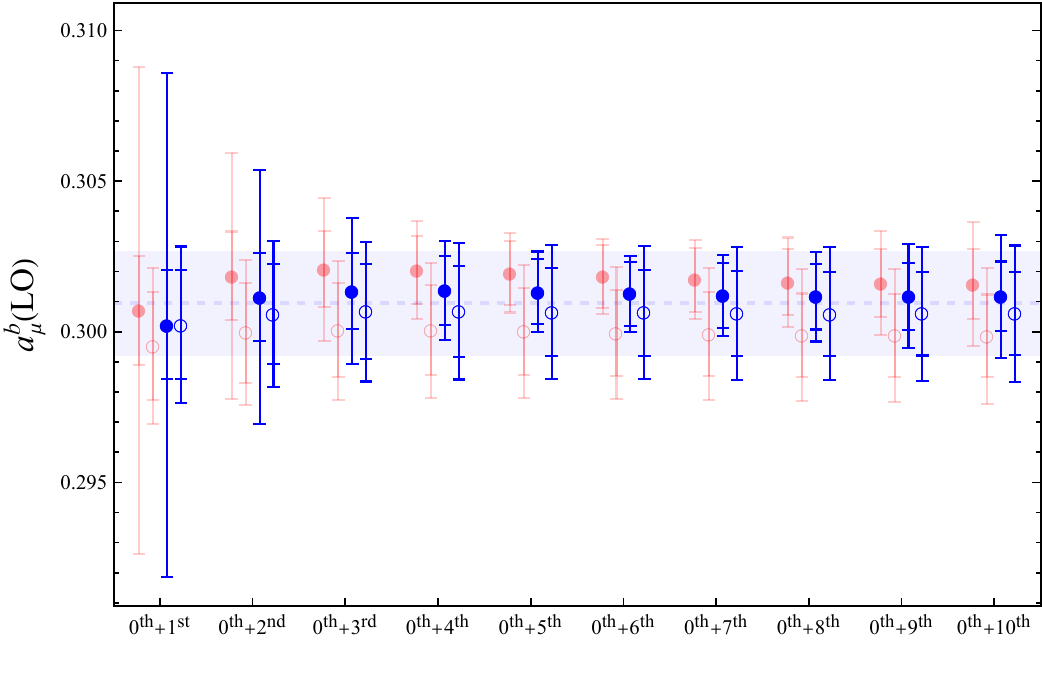}
\caption{$a_\mu^{b}({\rm LO})\times 10^{10}$ across different pairs and kernels}
\label{fig:amuLOb}
\end{subfigure}
\caption{(a) $\hat m_b(\hat m_b)$ and (b) $a_\mu^{b}({\rm LO})\times 10^{10}$, both as a function of the moment pair $(0^{\rm th}{+}n^{\rm th})$, obtained from the sum-rule self-consistency condition, Eq.~\eqref{eq:generalized_sumrule}, with the identity kernel (red) and with the kernel of the observable, $\hat K^{(2)}$ (blue). Panel (b) additionally distinguishes the theory-side (circles) and hadronic-side (empty points) evaluations of $a_\mu^{b}({\rm LO})$ for the mass shown in panel (a).}
\label{fig:mb_amuLOb}
\end{figure}

Figure~\ref{fig:mb} shows the resulting $\hat m_b$ as a function of the moment pair, for the two kernel choices; figure \ref{fig:amuLOb} shows how this same determination translates into $a_\mu^{b}({\rm LO})$, following exactly the same construction as the charm case, for both $K=1$ (red) and $K=\hat K^{(2)}$ (blue). The corresponding breakdown for the bottom sector is given in Table~\ref{tab:errorbudget_bottom}. As discussed together with the calibration procedure in Sec.~\ref{sec:expdata}, the treatment of the bottom-sector continuum has been further refined in this work, yielding a markedly more precise determination than in the charm sector.

In both sectors, two competing sources dominate the total uncertainty on $a_\mu^{q}({\rm LO})$: with increasing $n$, the calibration uncertainty from $\lambda_3^{q,\rm exp}$ decreases, while the truncation uncertainty increases, since higher moments are more sensitive to the region where the perturbative expansion is least reliable. These two sources combine to yield an optimal moment pair, beyond which neither further reduces the total uncertainty; this is reached at $0^{\rm th}{+}2^{\rm nd}$ for charm and $0^{\rm th}{+}6^{\rm th}$ for bottom, adopted as the default pairs in what follows.

Balancing this stability analysis against the residual moment-pair dependence, we adopt the aforementioned pair of moments as our default choice. Our final results, obtained from the sum rule weighted by the kernel of the observable, read
\begin{align}
\hat m_c(\hat m_c) &= (1267.1 \pm 6.8)~{\rm MeV}, \\
a_\mu^{c}({\rm LO}) &= (14.46 \pm 0.13)\times10^{-10}, \label{amucharm}
\end{align}
for the charm sector, and
\begin{align}
\hat m_b(\hat m_b) &= (4182.3 \pm 7.2)~{\rm MeV}, \\
a_\mu^{b}({\rm LO}) &= (0.3009 \pm 0.0017)\times10^{-10}, \label{amubottom}
\end{align}
for the bottom sector, with the corresponding correlation coefficients given in Tables~\ref{tab:corrmatrix_charm} and~\ref{tab:corrmatrix_bottom}.

These results in Eqs. \eqref{amucharm} and \eqref{amubottom} may be compared straightaway with a direct, non-correlated -with quark mass or hadronic correlations- result as the one extracted from the discussion presented in Ref.~\cite{Keshavarzi:2018mgv} (see the details in the dedicated Section \ref{sec:comparison}) for both charm and bottom. Such data driven results read $a_\mu^{c}({\rm LO})=14.44(23)\cdot 10^{-10}$ and $a_\mu^{b}({\rm LO})=0.30(2)\cdot 10^{-10}$, which implies our results represent an uncertainty reduction of $\sim 45\%$ and $\sim 90\%$ for charm and bottom, respectively.


\subsection{Calibrated uncertainty through experimental data}
\label{sec:expdata}

The determination of $\hat m_q$ and $\lambda_3^q$ presented above does not rely on experimental data for $R_q(s)$ beyond the resonance parameters of Table~\ref{tab:resonances}. A comparison with data for $R_q(s)$ in the threshold region is, however, used to calibrate the uncertainty of the determination. This is done by computing moments directly from available data and extracting the corresponding value of the continuum parameter, $\lambda_3^{q,\rm exp}$, which is compared with $\lambda_3^q$ obtained from the sum-rule self-consistency condition. Since the energy range covered by the experimental data in both charm and bottom sectors is finite ($\sim 1.1$GeV and $\sim 0.6$GeV, respectively), the extraction of $\lambda_3^{q,\rm exp}$ suffers from quark-hadron breaking effects hence we cannot expect this parameters would overlap statistically with $\lambda_3^q$. If it does, the result is reassessed and the uncertainty calibrated. We describe here the extraction of $\lambda_3^{q,\rm exp}$ for both quark sectors.

The experimental moments entering this calibration were already obtained, for $K=1$, in Refs.~\cite{Erler:2016atg,Erler:2022mzd} for the charm and bottom sectors, respectively. Since the generalized sum rules of Sec.~\ref{sec:generalized} additionally require the corresponding kernel-weighted moments, we reassess this extraction for the kernels of interest here.

Ref.~\cite{Erler:2016atg} obtained the charm moments by combining the available data collaboration by collaboration. To reassess the results, we instead perform the integration through a binning exploration, combining all available measurements within each energy bin. As expected, both methods yield very similar results. Since both methods agree, we adopt the moment obtained this way for $K=1$,
\begin{equation}
\int_{(2M_{D^0})^2}^{(4.8\,{\rm GeV})^2} \frac{ds}{s}\,R_c^{\rm cont}(s)\Big|_{\hat m_c} =
\mathcal{M}_0^{\rm exp} = 0.637(20)\, ,
\label{eq:M0exp_charm}
\end{equation}
consistent with Ref.~\cite{Erler:2016atg}, before extending the calculation to the kernel-weighted moments needed for this work.

\begin{figure}[t]
\centering
\includegraphics[width=\columnwidth]{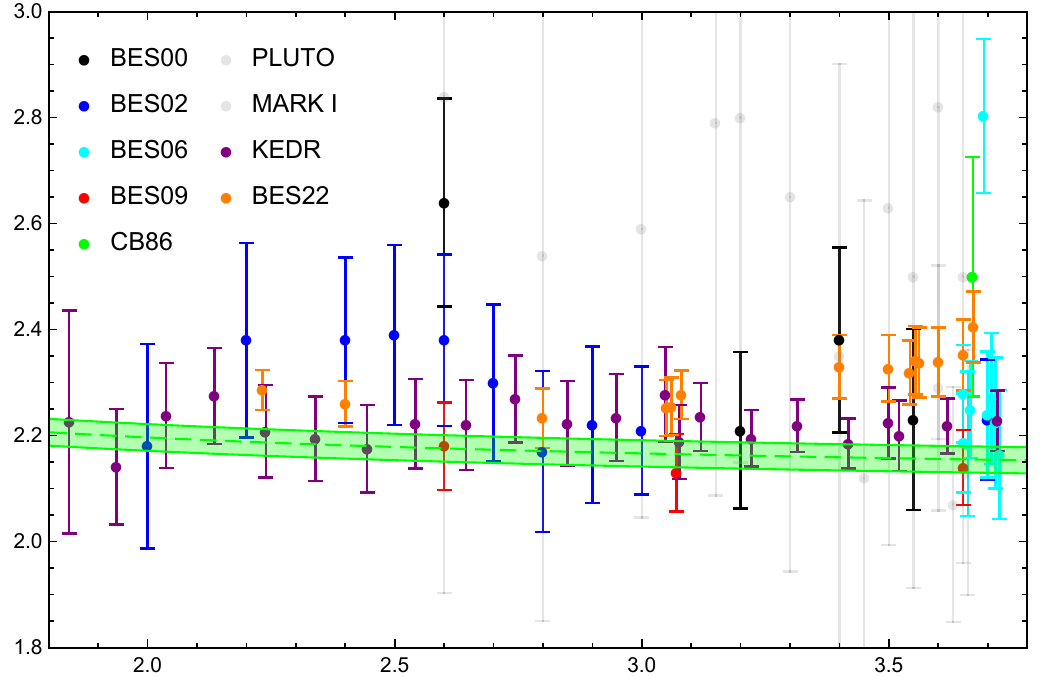}
\caption{\label{fig:kappa_fit}
Sub-threshold data ($\sqrt{s}<2M_{D^0}$) from all collaborations listed in Table~\ref{tab:data}, compared with the pQCD prediction for the light-quark background $R_{\rm bg}(s)$ rescaled by the global normalization factor $\kappa=0.998\pm0.011$ obtained from its combined fit.}
\end{figure}

Before integration, the non-resonant continuum pQCD background is subtracted, with its overall normalization $\kappa$ fixed by a fit to the data below the open-charm threshold. Table \ref{tab:data} details experimental collaborations and energy ranges used for the charm sector; $\kappa$ is understood as a global normalization of the data, not a modification of the theoretical prediction. Recent works have pointed out tensions between this pQCD-based subtraction and individual datasets, in particular the 2021 BES-III measurement~\cite{Boito:2025qfz,Kataev:2026gea}. We find that, while such tensions are indeed present when individual datasets are considered separately, they are no longer significant once the full set of collaborations is combined: performing the fit including all collaborations simultaneously, we obtain $\kappa=0.998\pm0.011$, compatible with unity, as shown in Fig.~\ref{fig:kappa_fit}. We nonetheless retain the uncertainty on $\kappa$ as an additional global systematic contribution to the experimental moment.

For the bottom case, we build on the combined BaBar~\cite{BaBar:2008cmq} and Belle~\cite{Belle:2015aea} dataset of Ref.~\cite{Erler:2022mzd}, which provides $R_b(s)$ corrected for vacuum polarization and QED radiative effects and with the non-resonant background already subtracted. As a cross-check, we reproduce the moment of Ref.~\cite{Dong:2020tdw} from this dataset for $K=1$ and find very good agreement using both. Thus, we adopt the value
\begin{equation}
\int_{(2M_B)^2}^{(11.20\,{\rm GeV})^2} \frac{ds}{s}\,R_b^{\rm cont}(s)\Big|_{\hat m_b} =
\mathcal{M}_0^{\rm exp} = 0.0446(11)\, .
\label{eq:M0exp_bottom}
\end{equation}
With this setup, we can easily extend the calculation to the kernels of interest.

\begin{table}[t]
\centering
\renewcommand{\arraystretch}{1.5}
\begin{tabular}{c @{\hspace{1.5cm}} c @{\hspace{1.5cm}} c}
\hline\hline
$K$ & Exp. moment & $\lambda_3^{q,\rm exp}$ \\
\hline
\textbf{Charm} & & $(\mathcal{A}_0,\mathcal{A}_2)^K$ \\
$1$            & $0.637(20)$   & $0.670(85)$ \\
$\hat K^{(2)}$ & $0.631(20)$   & $0.673(86)$ \\
\hline
\textbf{Bottom} & & $(\mathcal{A}_0,\mathcal{A}_6)^K$  \\
$1$            & $0.02005(45)$ & $0.691(45)$ \\
$\hat K^{(2)}$ & $0.02005(45)$ & $0.696(45)$ \\
\hline\hline
\end{tabular}
\caption{Experimental moment, for the default moment pair adopted in each sector, together with the resulting continuum-shape parameter $\lambda_3^{q,\rm exp}$, evaluated for $K=1$ and $K=\hat K^{(2)}$. For bottom, the experimental moment is obtained after subtracting the $\Upsilon(4S)$ and $\Upsilon(5S)$ contributions from the data, as described in the main text.}
\label{tab:lambda3exp}
\end{table}

In the bottom sector, Ref.~\cite{Erler:2022mzd} found it necessary to include the $\Upsilon(4S)$ and $\Upsilon(5S)$ states explicitly on top of the continuum ansatz, in order to stabilize the mass extraction against the non-trivial resonant structure just above threshold (Sec.~\ref{sec:sumrules}). For the calibration of $\lambda_3^{q,\rm exp}$, we instead compare the continuum component alone: we subtract the $\Upsilon(4S)$ and $\Upsilon(5S)$ contributions, parameterized as in Sec.~\ref{sec:sumrules}, from the data, and compare this continuum-only residual to $R_b^{\rm cont}(s)$ evaluated at the extracted $\hat m_b$. We conservatively associate to this subtraction an uncertainty covering $\pm3\Gamma_R$ around each resonance, roughly $90\%$ of its line shape, propagating the corresponding data uncertainty in that window into the continuum estimate. The reason for this choice is that the electronic widths of $\Upsilon(4S)$ and $\Upsilon(5S)$ entering $R_b^{\rm res}(s)$ are themselves extracted from the same cross-section data used for this calibration; comparing the full ansatz, resonances together with continuum, directly to the data would therefore count part of the resonance uncertainty twice, once already in the resonance parameters and once again in $\lambda_3^{q,\rm exp}$. Isolating the continuum avoids this, and ensures that the resulting discrepancy between $\lambda_3^q$ and $\lambda_3^{q,\rm exp}$ reflects genuinely missing non-perturbative effects, such as condensate contributions or residual duality violations, rather than a restatement of the resonance uncertainty already accounted for elsewhere. To extract the quark mass as was done in Ref.~\cite{Erler:2022mzd}, this detail is futile. Nonetheless, for $a_\mu^{b}({\rm LO,NLO})$ this may turn crucial, see Table \ref{tab:errorbudget_bottom} for the detailed error budget.

This procedure must not be applied to charm sector, where the $\psi(3770)$ and $\psi(4040)$ states, modeled as Dirac-$\delta$-functions in $R_c^{\rm res}(s)$, do not cover a physical data range and drop out of the comparison entirely: they can be regarded as part of the resonant component of the hadronic decomposition, Eq.~\eqref{eq:Rq_decomp}, rather than as an addition to the continuum that the data comparison must separately account for.

Table~\ref{tab:lambda3exp} collects, for the default moment pair adopted in each sector, $(\mathcal{M}_0,\mathcal{M}_2)$ for charm and $(\mathcal{M}_0,\mathcal{M}_6)$ for bottom, the resulting experimental moment (for bottom, with the $\Upsilon(4S)$ and $\Upsilon(5S)$ contributions subtracted as described above) together with its uncertainty, and the corresponding value of $\lambda_3^{q,\rm exp}$ with its uncertainty, evaluated for $K=1$ and $K=\hat K^{(2)}$.
A short summary is in order. In both charm and bottom cases, the extraction of the experimental zeroth (partial) moment is in very good agreement with either kernel used, and even better agreement is found when comparing the different $\lambda_3^{q,\rm exp}$ parameters. Compared to Refs.~\cite{Erler:2016atg,Erler:2022mzd} for charm and bottom $\lambda_3^{q,\rm exp}$ determinations, the results presented here are both smaller and more stable due to the extensions versos these previous works we introduced here (see Section \ref{sec:results}).

A concluding remark. The experimental comparison put forward in this section is not used as an additional determination of the central value, but only as an external calibration of the uncertainty associated with the continuum/duality modelling. The corresponding experimental information is therefore not included again as an independent constraint in the central extraction.

\subsection{Outlook: extending the framework beyond leading order}
\label{sec:nlo_outlook}

The construction of Sec.~\ref{sec:generalized} applies to any observable expressible as a kernel-weighted dispersive integral of the form of Eq.~\eqref{eq:amu_hvp}, and is not restricted to $\hat K^{(2)}$. We illustrate this here for the two NLO kernels, $\hat K^{(4a)}$ and $\hat K^{(4b)}$, showing that the same self-consistent extraction of $\hat m_q$ and the corresponding contribution carries over unchanged where the kernel allows it, and identifying where a genuine extension of the formalism is instead required.

\begin{figure}[t]
\centering
\includegraphics[width=\columnwidth]{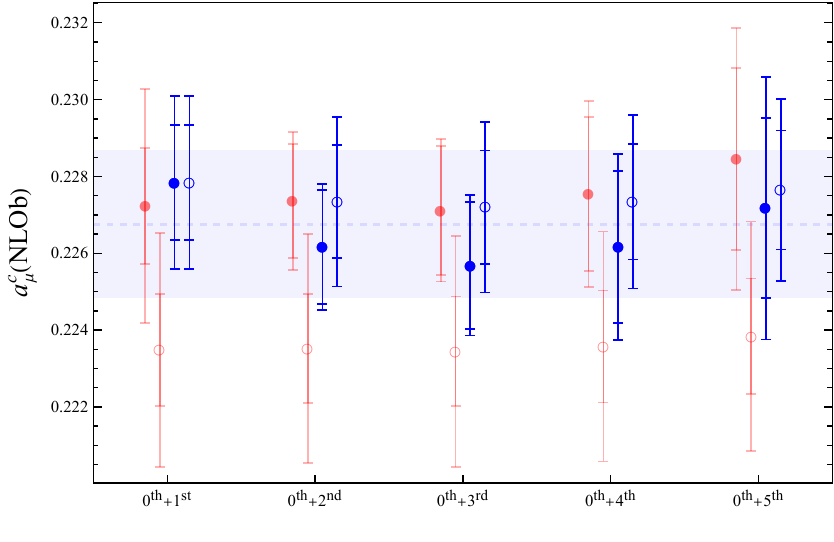}
\caption{\label{fig:amuNLObc}
$a_\mu^{c}({\rm NLO_b})\times 10^{10}$, both as a function of the moment pair $(0^{\rm th}{+}n^{\rm th})$, obtained from the sum-rule self-consistency condition, Eq.~\eqref{eq:generalized_sumrule}, with the identity kernel (red) and with the kernel of the observable, $\hat K^{(4b)}$ (blue), distinguishing the theory-side (circles) and hadronic-side (empty points) evaluations of $a_\mu^{c}({\rm NLO_b})$. Blue band reads $0.2268(19)$ with the obtained $\hat m_c(\hat m_c) = 1\,268.8(7.4)$ MeV.
}
\end{figure}

\begin{figure}[t]
\centering
\includegraphics[width=\columnwidth]{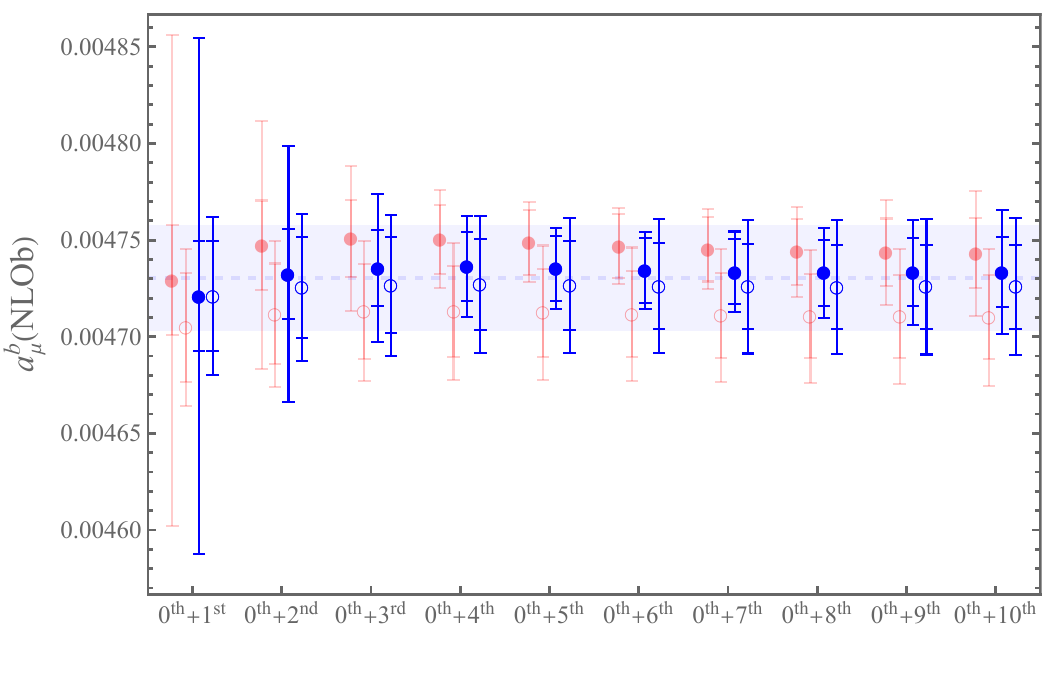}
\caption{\label{fig:amuNLObb}
$a_\mu^{b}({\rm NLO_b})\times 10^{10}$, both as a function of the moment pair $(0^{\rm th}{+}n^{\rm th})$, obtained from the sum-rule self-consistency condition, Eq.~\eqref{eq:generalized_sumrule}, with the identity kernel (red) and with the kernel of the observable, $\hat K^{(4b)}$ (blue), distinguishing the theory-side (circles) and hadronic-side (empty points) evaluations of $a_\mu^{b}({\rm NLO_b})$. Blue band reads $0.004730(27)$ with the obtained $\hat m_b(\hat m_b) = 4\,183.5(7.1)$ MeV.
}
\end{figure}

\begin{figure}[t]
\centering
\includegraphics[width=\columnwidth]{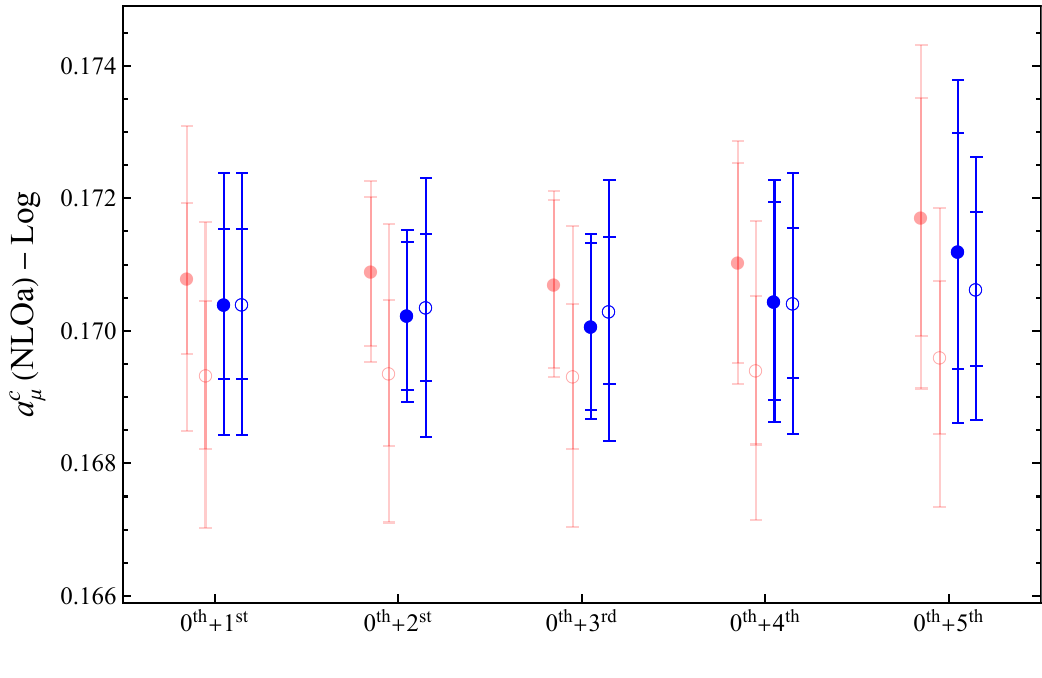}
\caption{\label{fig:amuNLOasub}
$(a_\mu^{c}({\rm NLO_a}) - \frac{23}{6} \ln z)\times 10^{10}$, both as a function of the moment pair $(0^{\rm th}{+}n^{\rm th})$, obtained from the sum-rule self-consistency condition, Eq.~\eqref{eq:generalized_sumrule}, with the identity kernel (red) and with the kernel of the observable, $\hat K^{(4a)}$ (blue), distinguishing the theory-side (circles) and hadronic-side (empty points) evaluations. 
}
\end{figure}

The kernel $\hat K^{(4b)}$ is structurally identical to $\hat K^{(2)}$ for the purposes of this construction: it is finite and directly integrable order by order in the high-energy expansion of Eq.~\eqref{eq:kernel_expansion}, with no additional subtraction needed beyond the one already introduced for the generalized zeroth moment in Eq.~\eqref{eq:A0_generalized}. The method of Sec.~\ref{sec:results} applies without modification, and Fig.~\ref{fig:amuNLObc} shows the resulting determination of $a_\mu^{q}({\rm NLOb})$, obtained exactly as for $a_\mu^{q}({\rm LO})$. 

The kernel $\hat K^{(4a)}$ presents an additional complication. Its high-energy expansion, Eq.~\eqref{eq:K4a}, contains a leading logarithmic term, $-\tfrac{23}{6}\ln z$, that is not suppressed by any power of $1/z$; as a consequence, the generalized zeroth moment $\mathcal{A}_0[\hat K^{(4a)}]$ built from this kernel is divergent, and a second, independent subtraction beyond the one already performed in Eq.~\eqref{eq:A0_generalized} would be required to define it. Constructing this second subtraction consistently, and matching it to the corresponding hadronic side, is beyond the scope of this work, as it would introduce further unknowns and thus artificially enlarge our uncertainty for a single subdominant NLO piece. Instead, as a validation of the method rather than an estimate of $a_\mu^{q}({\rm NLOa})$ itself, we apply it to the subtracted kernel $\hat K^{(4a)}_{\rm sub}(z) \equiv \hat K^{(4a)}(z) + \tfrac{23}{6}\ln z$, obtained by removing the divergent logarithm by hand; although this piece represents only a small fraction, roughly $20\%$, of the total $a_\mu^{q}({\rm NLOa})$ contribution, we find excellent agreement between the theory-side and hadronic-side descriptions of $a_\mu^{q}({\rm NLOa}_{\rm sub}) = a_\mu^{q}({\rm NLOa}) - \tfrac{23}{6}\ln z$, as shown in Fig.~\ref{fig:amuNLOasub} for the charm case. To provide an estimate of the full $a_\mu^{q}({\rm NLOa})$ contribution, no such obstruction applies to the higher moments: only the zeroth moment requires the further subtraction discussed above, which we consider beyond the scope of this work. We therefore take $\hat m_q$, together with its full correlated uncertainty, from the conventional sum-rule determination of Sec.~\ref{SubSec:ConstrainedSR}, and propagate it coherently into the theory-side evaluation of $a_\mu^{q}({\rm NLOa})$ alone. Since it is precisely the zeroth moment that is most sensitive to, and primarily responsible for fixing, the continuum shape parameter $\lambda_3^q$, we consider the theory-side evaluation to be the more reliable estimate of $a_\mu^{q}({\rm NLOa})$ in this particular case, in the absence of a self-consistent hadronic-side combination. We obtain $a_\mu^{c}({\rm NLOa}) = (-0.8006 \pm 0.0093)\times10^{-10}$ for the charm sector and $a_\mu^{b}({\rm NLOa}) = (-0.2295 \pm 0.0013)\times10^{-11}$ for the bottom sector, together with $a_\mu^{c}({\rm NLOb}) = (0.2268 \pm 0.0019)\times10^{-10}$ and $a_\mu^{b}({\rm NLOb}) = (0.04730 \pm 0.00027)\times10^{-11}$.

The remaining NLO contribution, $a_\mu^{q}({\rm NLOc})$, arising from diagrams with two hadronic insertions, is not itself a kernel-weighted dispersive integral of the form of Eq.~\eqref{eq:amu_hvp}, but a convolution of two hadronic spectral functions, and is left for future work.

\section{Comparison with previous works} 
\label{sec:comparison}

\begin{figure}[t]
\centering
\includegraphics[width=\columnwidth]{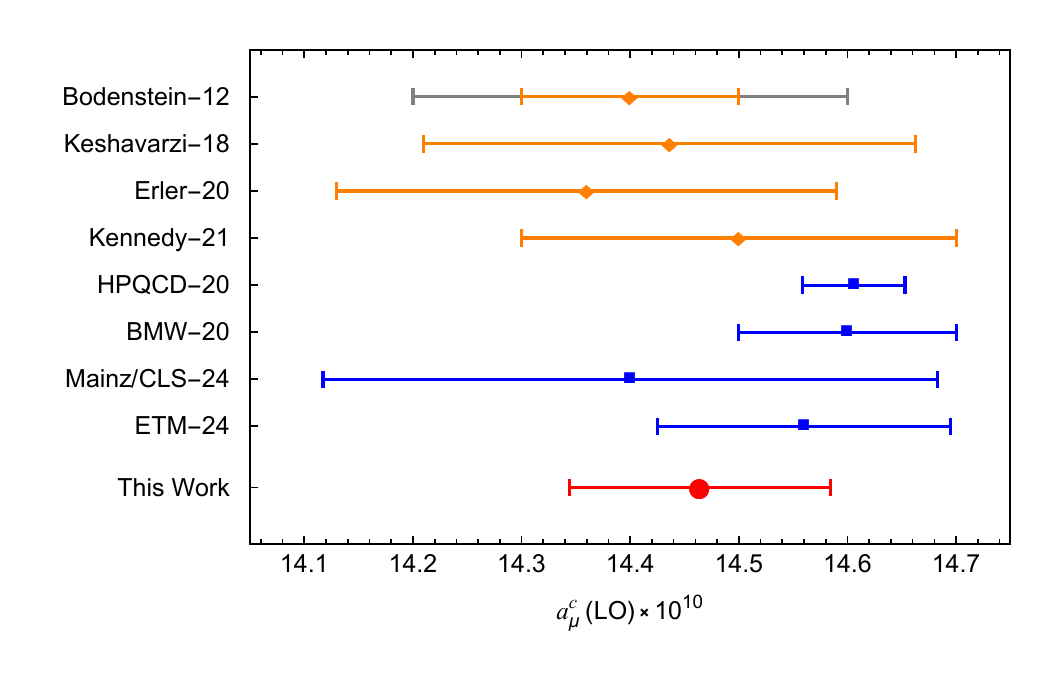}
\caption{\label{fig:comparison_LOc}
{Comparison of $a_\mu^{c}({\rm LO})$ with dispersive determinations (orange) and lattice determinations (blue). \textit{Bodenstein-12} result is shown with its original uncertainty (orange) \cite{Bodenstein:2011qy} and with the enlarged uncertainty reported in Ref.~\cite{Kennedy:2021ysp} (gray); \textit{Keshavarzi-18} result is our own determination from Ref.~\cite{Keshavarzi:2018mgv}, see the main text for details.}}
\end{figure}

Figure~\ref{fig:comparison_LOc} compares our determination of $a_\mu^{c}({\rm LO})$ with other dispersive evaluations (Bodenstain-12 from Ref.~\cite{Bodenstein:2011qy}, Keshavarzi-18 from~\cite{Keshavarzi:2018mgv}, Erler-20 from~\cite{Erler:2021bnl}) and with the recent reanalysis Kennedy-21 of Ref.~\cite{Kennedy:2021ysp}, together with lattice results Refs.~\cite{Hatton:2020qhk,Borsanyi:2020mff,ExtendedTwistedMass:2024nyi,Djukanovic:2024cmq}. The result Bodenstein-12 is shown with the uncertainty enlarged, in gray, to the value reported by Ref.~\cite{Kennedy:2021ysp}, who found that the original error estimate did not include the parametric uncertainty from $\alpha_s(\hat m_q)$ nor its correlation with $\hat m_q$, and obtained an uncertainty approximately twice as large when this was accounted for. The Keshavarzi-18 point is not a dedicated charm-sector determination; since Ref.~\cite{Keshavarzi:2018mgv} does not separate the charm and bottom contributions explicitly, we construct a naive estimate from Table~2 of that work, obtained by combining the $J/\psi$ and $\psi'$ resonance contributions directly, ascribing the pQCD tail above $\sqrt{s}=11.199$~GeV by the fractional-charge rule, and isolating the charm and bottom shares of the inclusive-channel contribution by integrating the same underlying data over the corresponding energy windows and subtracting the appropriate massless pQCD background. This procedure assigns approximately $14.2\%$ of the total inclusive contribution to charm and $0.1\%$ to bottom; we stress that this decomposition is our own estimate and not a result quoted in Ref.~\cite{Keshavarzi:2018mgv}. The remaining dispersive results are taken directly from the corresponding references, and the lattice results from Table~11 of the 2025 White Paper~\cite{Aliberti:2025beg}.

\begin{figure}[t]
\centering
\includegraphics[width=\columnwidth]{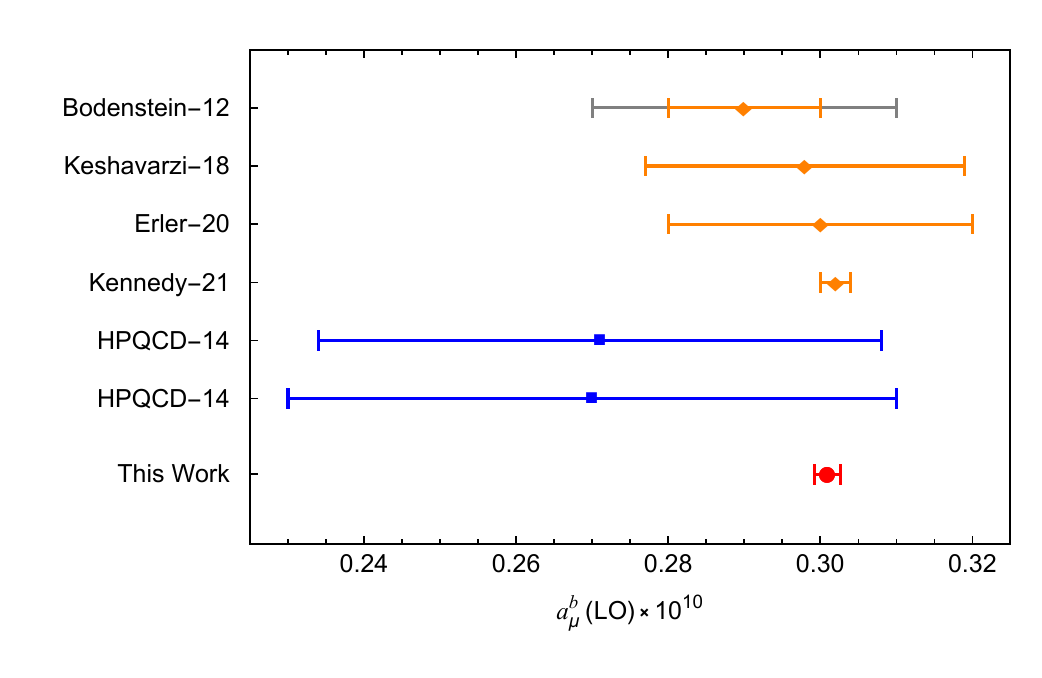}
\caption{\label{fig:comparison_LOb}
{Comparison of $a_\mu^{b}({\rm LO})$ with dispersive determinations (orange) and lattice determinations (blue). \textit{Bodenstein-12} result is shown with its original uncertainty (orange) from Ref. \cite{Bodenstein:2011qy} and with the enlarged uncertainty reported in Ref.~\cite{Kennedy:2021ysp} (gray); \textit{Keshavarzi-18} result is our own determination from Ref.~\cite{Keshavarzi:2018mgv}, see the main text for details.}
}
\end{figure}

Figure~\ref{fig:comparison_LOb} similarly compares our determination of $a_\mu^{b}({\rm LO})$ with the same dispersive evaluations, together with HPQCD lattice estimates from Refs.~\cite{Colquhoun:2014ica,FermilabLattice:2019ugu}.

\begin{figure}[t]
 \centering
 \includegraphics[width=\columnwidth]{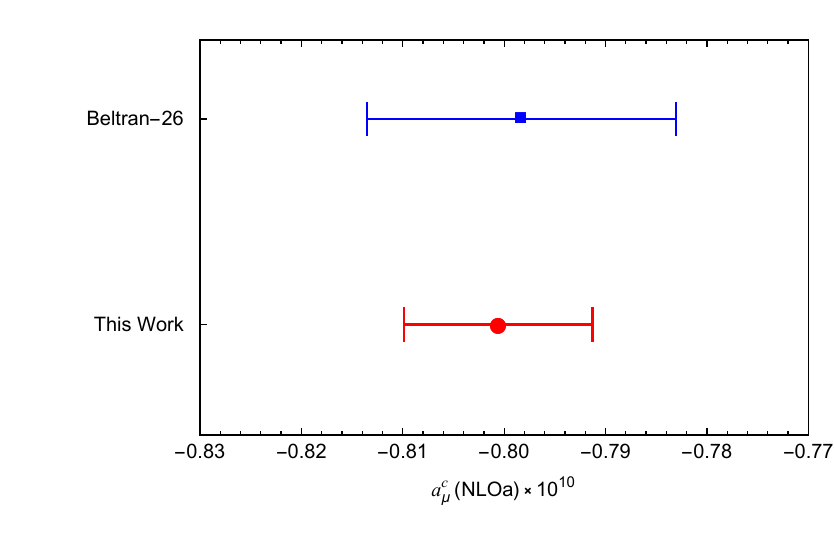}
 \caption{\label{fig:comparison_NLOa}
 Comparison of $a_\mu^{c}({\rm NLOa})$ with the lattice determination of Ref.~\cite{Beltran:2026ofp}.}
 \end{figure}

\begin{figure}[t]
\centering
\includegraphics[width=\columnwidth]{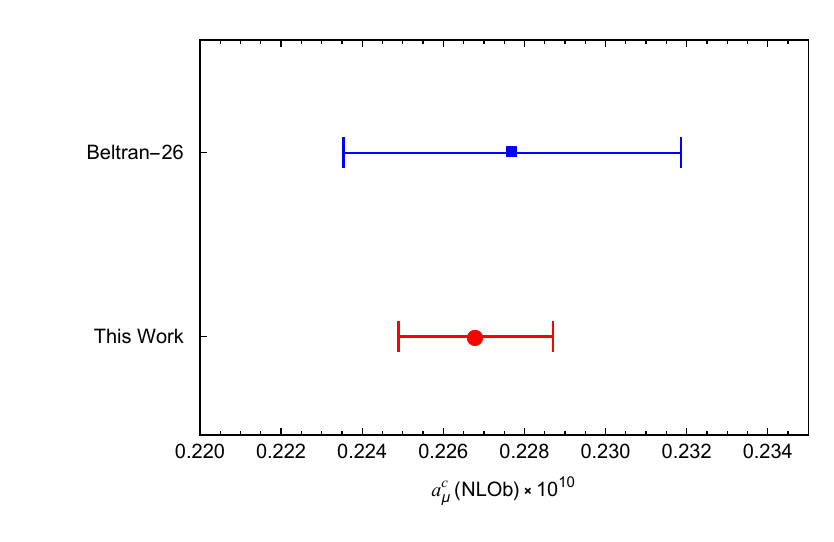}
\caption{\label{fig:comparison_NLOb}
Comparison of $a_\mu^{c}({\rm NLOb})$ with the lattice determination of Ref.~\cite{Beltran:2026ofp}.}
\end{figure}

Figures~\ref{fig:comparison_NLOa} and~\ref{fig:comparison_NLOb} show the analogous comparison for $a_\mu^{c}({\rm NLOa})$ and $a_\mu^{c}({\rm NLOb})$, respectively, together with the lattice determination of Ref.~\cite{Beltran:2026ofp}, the first available at this order.

Previous dispersive and lattice determinations of $a_\mu^{c,b}({\rm LO})$ treat the HVP contribution as the primary quantity to be determined, while conventional QCD sum-rule analyses use the heavy-quark mass as an external input when evaluating the corresponding HVP contribution. In contrast, our construction treats the mass and HVP contribution as correlated observables derived from the same spectral function and exploits this correlation as part of the determination itself, which induces a clear uncertainty reduction.

\section{Conclusions}
\label{sec:conclusions}

We have presented a determination of the heavy-quark contributions to the muon anomalous magnetic moment, $a_\mu^{c,b}$, obtained within a relativistic QCD sum-rule framework that determines the heavy-quark mass $\hat m_q$ and $a_\mu^{q}$ simultaneously, rather than treating one as an external input to the other. This unified treatment follows directly from the observation that both quantities are integrals over the same hadronic spectral function, differing only in their integration kernel, and that the sum-rule self-consistency condition used to fix $\hat m_q$ can equally be imposed using the physical HVP kernel in place of the ordinary power-law weight. The resulting generalized sum rules reduce, order by order in $\alpha_s$, to a finite perturbative expansion in the quark mass, exactly as for the ordinary moments, and require only a modest extension of the existing formalism to implement.

Using this framework, we obtain the mass parameters for charm and bottom quarks together with the correlated LO and NLO contributions to $a_\mu^{c,b}$ quoted in Sec.~\ref{sec:results}. The generalized sum rules make explicit that a value of $a_\mu^{q,(i)}$ obtained this way is only meaningful together with its correlated value of $\hat m_q$, and that the anticorrelation between the two, once properly accounted for, reduces the final uncertainty on $a_\mu^{q,(i)}$ relative to treating the mass as an independent external input. The residual spread between the theory-side and hadronic-side evaluations of $a_\mu^{q,(i)}$ at moment pairs other than the one used to fix $\hat m_q$ and $\lambda_3^q$ provides, in addition, a direct and physically meaningful diagnostic of residual theory/model dependence, including duality-violation and continuum-modeling systematics specific to the observable, one that is not available from the ordinary mass determination alone.

Two smaller extensions of the underlying continuum ansatz were found useful in reaching this result: the inclusion of the $\psi(3770)$ and $\psi(4040)$ resonances explicitly in the charm sector, in analogy with the treatment of the $\Upsilon(4S)$ and $\Upsilon(5S)$ states in the bottom sector, and the extension of the continuum mass corrections to $O(\hat m_q^4/s^2)$. Both improve the stability of the extracted mass across moment pairs, with the former relevant mainly for charm and the latter for bottom.

Our LO determination of $a_\mu^{c}$ is compatible with existing dispersive and lattice evaluations, and our NLO determinations, $a_\mu^{c}({\rm NLOa})$ and $a_\mu^{c}({\rm NLOb})$, provide the first comparison of this method against a lattice calculation at this order. For both quark sectors, the total uncertainty on $a_\mu^{q,(i)}$ is found to be dominated, beyond a certain moment pair, by the statistical uncertainty from the resonance parameters and, to a smaller extent, from $\hat\alpha_s(M_Z)$, rather than by the modeling of the continuum. Since this source of uncertainty is not tied to the assumptions of the continuum ansatz, it identifies a clear and model-independent direction along which future improvements in resonance spectroscopy could directly sharpen the heavy-quark HVP determination.

The generalized sum-rule construction introduced here is not specific to the heavy-quark HVP contribution and could in principle be applied to any observable expressible as a kernel-weighted dispersive integral over the same class of spectral functions. The heavy-quark HVP contribution studied here therefore provides a controlled proof of principle for applying this strategy to the broader HVP problem, with the longer-term aim of addressing the current tension between data-driven and lattice determinations of the light-quark HVP. The applicability of the construction, however, depends on the analytic and asymptotic properties of the kernel; the NLO $\hat{K}^{(4a)}$ case discussed below provides an explicit example where an additional subtraction is required. Extending the present analysis to other kernels, and to a more complete treatment of the correlation between the charm and bottom sectors, is left for future work.

\section*{Acknowledgements}

This work was supported in part by Deutsche Forschungsgemeinschaft (German Research Foundation, DFG) through the research unit FOR 5327 “Photon-photon interactions in the Standard Model and beyond exploiting the discovery potential from MESA to the LHC” (Project No. 458854507) and through the Cluster of Excellence “Precision Physics, Fundamental Interactions and Structure of Matter” (PRISMA+ EXC 2118/1), funded within the German Excellence strategy (Project No. 390831469), and in part by the Ministerio de Ciencia e Innovación under grant PID2020-112965GB-I00, by the Secretaria d’Universitats i Recerca del Departament d’Empresa i Coneixement de la Generalitat de Catalunya under grant 2021 SGR 00649, and by the Spanish Ministry of Science and Innovation (MICINN) through the State Research Agency under the Severo Ochoa Centres of Excellence Programme 2025–2029 (CEX2024-001441-S). IFAE is partially funded by the CERCA program of the Generalitat de Catalunya.


\appendix

\section{High-energy expansion of QED kernels}
\label{app:kernel_expansion}

This appendix contains the high-energy expansions of the QED kernels $\hat{K}^{(2)}(z)$, $\hat{K}^{(4a)}(z)$, and $\hat{K}^{(4b)}(z)$ used in the present work.

\begin{widetext}
\begin{align}
\label{eq:K2}
  \hat{K}^{(2)}(z)
    &= 1
      +\frac{1}{z}\!\left(\frac{25}{4}-3\ln z\right)
      +\frac{1}{z^2}\!\left(\frac{291}{10}-18\ln z\right)
      +\frac{1}{z^3}\!\left(\frac{624}{5}-84\ln z\right)
\end{align}
\begin{align}
\label{eq:K4a}
  \hat{K}^{(4a)}(z)
    &= \frac{223}{9}-12\zeta(2)\boldsymbol{-\frac{23}{6}\ln z}
  \nonumber\\
    &\quad+\frac{1}{z}\!\left(\frac{8785}{192}-\frac{111}{4}\zeta(2)
      -\frac{367}{36}\ln z+\frac{19}{24}\ln^2\!z\right)
  \nonumber\\
    &\quad+\frac{1}{z^2}\!\left(\frac{13072841}{72000}-\frac{2649}{20}\zeta(2)
      -\frac{10079}{600}\ln z+\frac{423}{40}\ln^2\!z\right)
  \nonumber\\
    &\quad+\frac{1}{z^3}\!\left(\frac{6104109}{8000}-\frac{11709}{20}\zeta(2)
      -\frac{6517}{300}\ln z+\frac{2883}{40}\ln^2\!z\right)
\end{align}
\begin{align}\label{eq:K4basy}
  \hat{K}^{(4b)}(z)
    &= -\frac{1}{3}+\frac{2}{3}\ln\!\frac{m_\mu^2}{m_f^2}
  \nonumber\\
    &\quad+\frac{1}{z}\!\left(-\frac{55}{8}+\frac{\pi^2}{3}
      +\frac{25}{6}\ln\!\frac{m_\mu^2}{m_f^2}
      +\frac{10}{3}\ln z
      -2\ln\!\frac{m_\mu^2}{m_f^2}\ln z-\ln^2\!z\right)
  \nonumber\\
    &\quad+\frac{1}{z^2}\!\left(-\frac{11299}{300}+2\pi^2
      +\frac{97}{5}\ln\!\frac{m_\mu^2}{m_f^2}
      +20\ln z
      -12\ln\!\frac{m_\mu^2}{m_f^2}\ln z-6\ln^2\!z\right)
  \nonumber\\
    &\quad+\frac{1}{z^3}\!\left(-\frac{12838}{75}+\frac{28}{3}\pi^2
      +\frac{416}{5}\ln\!\frac{m_\mu^2}{m_f^2}
      +\frac{280}{3}\ln z
      -56\ln\!\frac{m_\mu^2}{m_f^2}\ln z-28\ln^2\!z\right).
\end{align}
\end{widetext}

where $m_f$ is the mass of the fermion in the loop~\cite{Krause:1996rf}.


\section{Numerical coefficients $C_{n,k}^{(i)}$}
\label{app:Cnk}

Table \ref{tab:Cnk} collects the numerical coefficients $C_{n,k}^{(i)}$ for either the charm sector ($n_l=3$) and bottom sector ($n_l=4$), respectively. These numerical results are obtained after solving Eq.\eqref{eq:Cnr_def}.

\begin{table*}[t]
\caption{Numerical coefficients $C_{n,k}^{(i)}$ for $k=0,1,2$.
The coefficients $C_{n,k}^{(0)}$ and $C_{n,k}^{(1)}$ are independent
of $n_l$, while $C_{n,k}^{(2)}$ and $C_{n,k}^{(3)}$ depend on $n_l$.
The coefficients for $i=0,1,2$ are taken from \cite{Chetyrkin:1997mb,Maier:2007yn},
and for $i=3$ from \cite{Kiyo:2009gb}, later recomputed in \cite{Greynat:2011zp}
with more conservative uncertainties. The $r=1,2$ coefficients are obtained
from a reconstruction of the correlator following Refs.~\cite{Greynat:2010kx,Greynat:2011zp}.
\label{tab:Cnk}}
\centering
\bgroup
\sisetup{retain-zero-uncertainty = true}
\def\arraystretch{1.25}
\setlength{\tabcolsep}{8pt}
\begin{tabular}{|c|S[table-format=1.4]S[table-format=1.4]|S[table-format=-2.4(2)]S[parse-numbers=false]|S[table-format=-2.4(2)]S[parse-numbers=false]|}
\hline
& & & \multicolumn{2}{c|}{$n_l=3$}
    & \multicolumn{2}{c|}{$n_l=4$} \\
{$n$}
  & {$C_{n,0}^{(0)}$} & {$C_F C_{n,0}^{(1)}$}
  & {$C_{n,0}^{(2)}$} & {$C_{n,0}^{(3)}$}
  & {$C_{n,0}^{(2)}$} & {$C_{n,0}^{(3)}$} \\
\hline
$1$  & 1.0667 & 5.3992 & 33.2201 & 366.1750  & 31.6608 & 308.0190  \\
$2$  & 0.4571 & 3.5476 & 32.4404 & 381.5090  & 30.9911 & 330.5840  \\
$3$  & 0.2709 & 2.6866 & 29.7909 & 385.2330  & 28.5271 & 338.7070  \\
$4$  & 0.1847 & 2.1733 & 27.4082 & 382.7(5) & 26.2905 & 339.7(5) \\
$5$  & 0.1364 & 1.8293 & 25.4068 & 378.0(1.2)    & 24.4106 & 337.7(1.2)  \\
$6$  & 0.1061 & 1.5816 & 23.7378 & 372.5(1.8)    & 22.8356 & 334.5(1.8)  \\
$7$  & 0.0856 & 1.3942 & 22.3285 & 367.0(2.3)    & 21.5028 & 330.9(2.3)  \\
$8$  & 0.0709 & 1.2474 & 21.1231 & 362.5(2.7)    & 20.3609 & 327.2(2.7)  \\
$9$  & 0.0601 & 1.1292 & 20.0813 & 356.4(3.1)    & 19.3727 & 323.5(3.1)  \\
$10$ & 0.0517 & 1.0317 & 19.1647 & 352.6(3.4)    & 18.5019 & 320.0(3.4)    \\
\hline
\hline
& & & \multicolumn{2}{c|}{$n_l=3$}
    & \multicolumn{2}{c|}{$n_l=4$} \\
{$n$}
  & {$C_{n,1}^{(0)}$} & {$C_F C_{n,1}^{(1)}$}
  & {$C_{n,1}^{(2)}$} & {$C_{n,1}^{(3)}$}
  & {$C_{n,1}^{(2)}$} & {$C_{n,1}^{(3)}$} \\
\hline
$1$  & 1.2591 & 3.0333 & -3.9412(254) & -11.0(11.0) & -3.7690(254) & -23.7(23.7)   \\
$2$  & 0.2893 & 1.1618 & 2.5306(76)  & -9.2(9.2) & 2.3356(76)  & -14.3(14.3)   \\
$3$  & 0.1187 & 0.6413 & 2.5822(39)  & 0.4(4) & 2.4135(39)  & -3.5(3.5)  \\
$4$  & 0.0622 & 0.4115 & 2.1920(24)  & 4.0(4.0) & 2.0595(24)  & 0.9(9)  \\
$5$  & 0.0374 & 0.2879 & 1.8264(17)  & 5.3(5.3) & 1.7217(17)  & 2.7(2.7)   \\
$6$  & 0.0245 & 0.2132 & 1.5329(12)  & 5.6(5.6) & 1.4485(12)  & 3.5(3.5)   \\
$7$  & 0.0171 & 0.1646 & 1.3026(10)  & 5.5(5.5) & 1.2331(10)  & 3.7(3.7)   \\
$8$  & 0.0125 & 0.1310 & 1.1209(8)   & 5.3(5.3) & 1.0627(8)   & 3.7(3.7)   \\
$9$  & 0.0095 & 0.0126 & 0.9757(6)   & 5.0(5.0) & 0.9262(6)   & 3.6(3.6)   \\
$10$ & 0.0074 & 0.0098 & 0.8582(5)   & 4.7(4.7) & 0.8155(5)   & 3.5(3.5)   \\
\hline
\hline
& & & \multicolumn{2}{c|}{$n_l=3$}
    & \multicolumn{2}{c|}{$n_l=4$} \\
{$n$}
  & {$C_{n,2}^{(0)}$} & {$C_F C_{n,2}^{(1)}$}
  & {$C_{n,2}^{(2)}$} & {$C_{n,2}^{(3)}$}
  & {$C_{n,2}^{(2)}$} & {$C_{n,2}^{(3)}$} \\
\hline
$1$  & 2.6217 & 4.1593 & -22.5852(458) & 83.5(83.5) & -21.0964(458) & 55.6(55.6)    \\
$2$  & 0.3134 & 0.8277 & -0.8880(64)  & -14.3(14.3) & -0.8767(64)  & -16.1(16.1)   \\
$3$  & 0.0881 & 0.3227 & 0.3334(21)   & -5.8(5.8) & 0.2945(21)   & -6.7(6.7)  \\
$4$  & 0.0353 & 0.1621 & 0.3970(10)   & -2.1(2.1) & 0.3648(10)   & -2.7(2.7)  \\
$5$  & 0.0172 & 0.0936 & 0.3296(5)    & -0.7(0.7) & 0.3060(5)    & -1.1(1.1)  \\
$6$  & 0.0095 & 0.0592 & 0.2595(3)    & -0.1(0.1) & 0.2422(3)    & -0.4(0.4) \\
$7$  & 0.0057 & 0.0399 & 0.2038(2)    & 0.2(0.2) & 0.1908(2)    & -0.1(0.1)\\
$8$  & 0.0037 & 0.0282 & 0.1617(2)    & 0.3(0.3) & 0.1518(2)    & 0.1(0.1) \\
$9$  & 0.0025 & 0.0033 & 0.1301(0)    & 0.3(0.3) & 0.1224(1)    & 0.1(0.1)  \\
$10$ & 0.0018 & 0.0023 & 0.1061(1)    & 0.3(0.3) & 0.1000(1)    & 0.2(0.2)  \\
\hline
\end{tabular}
\egroup
\end{table*}


\section{Correlation matrices}\label{App:CorrelationMatrices}


Tables \ref{tab:corrmatrix_charm} and \ref{tab:corrmatrix_bottom} collect the correlation matrices mentioned at the end of Sec. \ref{SubSec:ConstrainedSR}.

\begin{table*}[t]
\caption{\label{tab:corrmatrix_charm}
Correlation matrix among the parameters entering the charm-sector
determination and the resulting mass and $a_\mu^{c,\rm LO}$
contributions, for the default moment pair $(\mathcal{M}_0,\mathcal{M}_2)$.
Resonance labels refer to the electronic width $\Gamma_R^{e}$, the
only resonance parameter entering the error budget. Correlations
among the input parameters (bold) are set by construction: the
electronic widths of $J/\psi$ and $\psi(2S)$ are taken to be $50\%$
correlated, while $\alpha_Z$, $C_G$, and all other resonance pairs are
treated as independent. Cell shading encodes correlation strength, red
for positive, blue for negative, saturating at $|\rho|=1$.}
\centering
\renewcommand{\arraystretch}{1.2}
\setlength{\tabcolsep}{3pt}
\footnotesize
\resizebox{\textwidth}{!}{%
\begin{tabular}{|l|cccccc|cccc|}
\hline
 & $\alpha_Z$ & $J/\psi$ & $\psi(2S)$ & $\psi(3770)$ & $\psi(4040)$ & $C_G$ & $\hat m_c$ & $\lambda_3^{c}$ & $a_\mu^{c,\rm th,LO}$ & $a_\mu^{c,\rm had,LO}$ \\
\hline
$\alpha_Z$ & \cellcolor{red!60}\textbf{1.00} & \textbf{0.00} & \textbf{0.00} & \textbf{0.00} & \textbf{0.00} & \textbf{0.00} & \cellcolor{red!10}0.17 & \cellcolor{red!15}0.24 & \cellcolor{red!44}0.73 & \cellcolor{red!28}0.46 \\
$J/\psi$ & \textbf{0.00} & \cellcolor{red!60}\textbf{1.00} & \cellcolor{red!30}\textbf{0.50} & \textbf{0.00} & \textbf{0.00} & \textbf{0.00} & \cellcolor{blue!41}-0.68 & \cellcolor{blue!40}-0.66 & \cellcolor{red!42}0.70 & \cellcolor{red!56}0.93 \\
$\psi(2S)$ & \textbf{0.00} & \cellcolor{red!30}\textbf{0.50} & \cellcolor{red!60}\textbf{1.00} & \textbf{0.00} & \textbf{0.00} & \textbf{0.00} & \cellcolor{blue!15}-0.25 & \cellcolor{blue!35}-0.58 & \cellcolor{red!11}0.18 & \cellcolor{red!25}0.42 \\
$\psi(3770)$ & \textbf{0.00} & \textbf{0.00} & \textbf{0.00} & \cellcolor{red!60}\textbf{1.00} & \textbf{0.00} & \textbf{0.00} & \cellcolor{red!3}0.05 & \cellcolor{blue!6}-0.10 & \cellcolor{blue!3}-0.06 & \cellcolor{blue!2}-0.03 \\
$\psi(4040)$ & \textbf{0.00} & \textbf{0.00} & \textbf{0.00} & \textbf{0.00} & \cellcolor{red!60}\textbf{1.00} & \textbf{0.00} & \cellcolor{blue!5}-0.08 & \cellcolor{blue!33}-0.54 & \cellcolor{red!2}0.03 & \cellcolor{red!9}0.15 \\
$C_G$ & \textbf{0.00} & \textbf{0.00} & \textbf{0.00} & \textbf{0.00} & \textbf{0.00} & \cellcolor{red!60}\textbf{1.00} & \cellcolor{blue!41}-0.68 & \cellcolor{red!22}0.37 & \cellcolor{red!27}0.45 & \cellcolor{red!13}0.22 \\
\hline
$\hat m_c$ & \cellcolor{red!10}0.17 & \cellcolor{blue!41}-0.68 & \cellcolor{blue!15}-0.25 & \cellcolor{red!3}0.05 & \cellcolor{blue!5}-0.08 & \cellcolor{blue!41}-0.68 & \cellcolor{red!60}1.00 & \cellcolor{red!20}0.33 & \cellcolor{blue!32}-0.53 & \cellcolor{blue!38}-0.64 \\
$\lambda_3^{c}$ & \cellcolor{red!15}0.24 & \cellcolor{blue!40}-0.66 & \cellcolor{blue!35}-0.58 & \cellcolor{blue!6}-0.10 & \cellcolor{blue!33}-0.54 & \cellcolor{red!22}0.37 & \cellcolor{red!20}0.33 & \cellcolor{red!60}1.00 & \cellcolor{blue!8}-0.14 & \cellcolor{blue!32}-0.54 \\
$a_\mu^{c,\rm th,LO}$ & \cellcolor{red!44}0.73 & \cellcolor{red!42}0.70 & \cellcolor{red!11}0.18 & \cellcolor{blue!3}-0.06 & \cellcolor{red!2}0.03 & \cellcolor{red!27}0.45 & \cellcolor{blue!32}-0.53 & \cellcolor{blue!8}-0.14 & \cellcolor{red!60}1.00 & \cellcolor{red!54}0.90 \\
$a_\mu^{c,\rm had,LO}$ & \cellcolor{red!28}0.46 & \cellcolor{red!56}0.93 & \cellcolor{red!25}0.42 & \cellcolor{blue!2}-0.03 & \cellcolor{red!9}0.15 & \cellcolor{red!13}0.22 & \cellcolor{blue!38}-0.64 & \cellcolor{blue!32}-0.54 & \cellcolor{red!54}0.90 & \cellcolor{red!60}1.00 \\
\hline
\end{tabular}}
\end{table*}

\begin{table*}[t]
\caption{\label{tab:corrmatrix_bottom}
Correlation matrix among the parameters entering the bottom-sector
determination and the resulting mass and $a_\mu^{b,\rm LO}$
contributions, for the default moment pair $(\mathcal{M}_0,\mathcal{M}_6)$.
Resonance labels refer to the electronic width $\Gamma_R^{e}$, the
only resonance parameter entering the error budget. Correlations
among the input parameters (bold) are set by construction: the
electronic widths of $\Upsilon(1S)$, $\Upsilon(2S)$, and $\Upsilon(3S)$
are taken to be pairwise $50\%$ correlated, while $\alpha_Z$, $C_G$,
$\Upsilon(4S)$, and $\Upsilon(5S)$ are treated as independent of all
other inputs. Cell shading encodes correlation strength, red for
positive, blue for negative, saturating at $|\rho|=1$.}
\centering
\renewcommand{\arraystretch}{1.2}
\setlength{\tabcolsep}{3pt}
\scriptsize
\resizebox{\textwidth}{!}{%
\begin{tabular}{|l|ccccccc|cccc|}
\hline
 & $\alpha_Z$ & $\Upsilon(1S)$ & $\Upsilon(2S)$ & $\Upsilon(3S)$ & $\Upsilon(4S)$ & $\Upsilon(5S)$ & $C_G$ & $\hat m_b$ & $\lambda_3^{b}$ & $a_\mu^{b,\rm th,LO}$ & $a_\mu^{b,\rm had,LO}$ \\
\hline
$\alpha_Z$ & \cellcolor{red!60}\textbf{1.00} & \textbf{0.00} & \textbf{0.00} & \textbf{0.00} & \textbf{0.00} & \textbf{0.00} & \textbf{0.00} & \cellcolor{blue!21}-0.35 & \cellcolor{red!6}0.09 & \cellcolor{red!51}0.86 & \cellcolor{red!28}0.47 \\
$\Upsilon(1S)$ & \textbf{0.00} & \cellcolor{red!60}\textbf{1.00} & \cellcolor{red!30}\textbf{0.50} & \cellcolor{red!30}\textbf{0.50} & \textbf{0.00} & \textbf{0.00} & \textbf{0.00} & \cellcolor{blue!53}-0.88 & \cellcolor{blue!18}-0.30 & \cellcolor{red!33}0.55 & \cellcolor{red!36}0.60 \\
$\Upsilon(2S)$ & \textbf{0.00} & \cellcolor{red!30}\textbf{0.50} & \cellcolor{red!60}\textbf{1.00} & \cellcolor{red!30}\textbf{0.50} & \textbf{0.00} & \textbf{0.00} & \textbf{0.00} & \cellcolor{blue!33}-0.56 & \cellcolor{blue!13}-0.22 & \cellcolor{red!20}0.33 & \cellcolor{red!24}0.40 \\
$\Upsilon(3S)$ & \textbf{0.00} & \cellcolor{red!30}\textbf{0.50} & \cellcolor{red!30}\textbf{0.50} & \cellcolor{red!60}\textbf{1.00} & \textbf{0.00} & \textbf{0.00} & \textbf{0.00} & \cellcolor{blue!41}-0.69 & \cellcolor{blue!16}-0.26 & \cellcolor{red!24}0.40 & \cellcolor{red!29}0.48 \\
$\Upsilon(4S)$ & \textbf{0.00} & \textbf{0.00} & \textbf{0.00} & \textbf{0.00} & \cellcolor{red!60}\textbf{1.00} & \textbf{0.00} & \textbf{0.00} & \cellcolor{blue!2}-0.03 & \cellcolor{blue!21}-0.34 & \cellcolor{red!0}0.01 & \cellcolor{red!13}0.21 \\
$\Upsilon(5S)$ & \textbf{0.00} & \textbf{0.00} & \textbf{0.00} & \textbf{0.00} & \textbf{0.00} & \cellcolor{red!60}\textbf{1.00} & \textbf{0.00} & \cellcolor{blue!23}-0.39 & \cellcolor{blue!53}-0.89 & \cellcolor{red!15}0.25 & \cellcolor{red!40}0.67 \\
$C_G$ & \textbf{0.00} & \textbf{0.00} & \textbf{0.00} & \textbf{0.00} & \textbf{0.00} & \textbf{0.00} & \cellcolor{red!60}\textbf{1.00} & \cellcolor{blue!0}-0.01 & \cellcolor{blue!5}-0.08 & \cellcolor{red!5}0.08 & \cellcolor{red!4}0.07 \\
\hline
$\hat m_b$ & \cellcolor{blue!21}-0.35 & \cellcolor{blue!53}-0.88 & \cellcolor{blue!33}-0.56 & \cellcolor{blue!41}-0.69 & \cellcolor{blue!2}-0.03 & \cellcolor{blue!23}-0.39 & \cellcolor{blue!0}-0.01 & \cellcolor{red!60}1.00 & \cellcolor{red!34}0.57 & \cellcolor{blue!47}-0.79 & \cellcolor{blue!54}-0.90 \\
$\lambda_3^{b}$ & \cellcolor{red!6}0.09 & \cellcolor{blue!18}-0.30 & \cellcolor{blue!13}-0.22 & \cellcolor{blue!16}-0.26 & \cellcolor{blue!21}-0.34 & \cellcolor{blue!53}-0.89 & \cellcolor{blue!5}-0.08 & \cellcolor{red!34}0.57 & \cellcolor{red!60}1.00 & \cellcolor{blue!15}-0.25 & \cellcolor{blue!46}-0.77 \\
$a_\mu^{b,\rm th,LO}$ & \cellcolor{red!51}0.86 & \cellcolor{red!33}0.55 & \cellcolor{red!20}0.33 & \cellcolor{red!24}0.40 & \cellcolor{red!0}0.01 & \cellcolor{red!15}0.25 & \cellcolor{red!5}0.08 & \cellcolor{blue!47}-0.79 & \cellcolor{blue!15}-0.25 & \cellcolor{red!60}1.00 & \cellcolor{red!48}0.81 \\
$a_\mu^{b,\rm had,LO}$ & \cellcolor{red!28}0.47 & \cellcolor{red!36}0.60 & \cellcolor{red!24}0.40 & \cellcolor{red!29}0.48 & \cellcolor{red!13}0.21 & \cellcolor{red!40}0.67 & \cellcolor{red!4}0.07 & \cellcolor{blue!54}-0.90 & \cellcolor{blue!46}-0.77 & \cellcolor{red!48}0.81 & \cellcolor{red!60}1.00 \\
\hline
\end{tabular}}
\end{table*}

\begin{table*}[t]
\caption{\label{tab:corrmatrix_combined}
Correlation matrix between the charm- and bottom-sector output
quantities, obtained from the same Monte Carlo ensemble in which the
shared inputs ($\alpha_Z$, $C_G$) are drawn consistently across both
flavour determinations. Cell shading encodes correlation strength, red
for positive, blue for negative, saturating at $|\rho|=1$.}
\centering
\renewcommand{\arraystretch}{1.2}
\setlength{\tabcolsep}{3pt}
\footnotesize
\resizebox{\textwidth}{!}{%
\begin{tabular}{|l|cccc|cccc|}
\hline
 & $\hat m_c$ & $\lambda_3^{c}$ & $a_\mu^{c,\rm th,LO}$ & $a_\mu^{c,\rm had,LO}$ & $\hat m_b$ & $\lambda_3^{b}$ & $a_\mu^{b,\rm th,LO}$ & $a_\mu^{b,\rm had,LO}$ \\
\hline
$\hat m_c$ & \cellcolor{red!60}1.00 & \cellcolor{red!20}0.33 & \cellcolor{blue!32}-0.53 & \cellcolor{blue!38}-0.64 & \cellcolor{blue!10}-0.16 & \cellcolor{blue!2}-0.04 & \cellcolor{red!12}0.21 & \cellcolor{red!10}0.17 \\
$\lambda_3^{c}$ & \cellcolor{red!20}0.33 & \cellcolor{red!60}1.00 & \cellcolor{blue!8}-0.14 & \cellcolor{blue!32}-0.54 & \cellcolor{blue!8}-0.13 & \cellcolor{blue!6}-0.09 & \cellcolor{red!14}0.23 & \cellcolor{red!11}0.19 \\
$a_\mu^{c,\rm th,LO}$ & \cellcolor{blue!32}-0.53 & \cellcolor{blue!8}-0.14 & \cellcolor{red!60}1.00 & \cellcolor{red!54}0.90 & \cellcolor{blue!11}-0.18 & \cellcolor{red!8}0.13 & \cellcolor{red!35}0.58 & \cellcolor{red!16}0.27 \\
$a_\mu^{c,\rm had,LO}$ & \cellcolor{blue!38}-0.64 & \cellcolor{blue!32}-0.54 & \cellcolor{red!54}0.90 & \cellcolor{red!60}1.00 & \cellcolor{blue!4}-0.06 & \cellcolor{red!10}0.17 & \cellcolor{red!20}0.34 & \cellcolor{red!6}0.10 \\
\hline
$\hat m_b$ & \cellcolor{blue!10}-0.16 & \cellcolor{blue!8}-0.13 & \cellcolor{blue!11}-0.18 & \cellcolor{blue!4}-0.06 & \cellcolor{red!60}1.00 & \cellcolor{red!34}0.57 & \cellcolor{blue!47}-0.79 & \cellcolor{blue!54}-0.90 \\
$\lambda_3^{b}$ & \cellcolor{blue!2}-0.04 & \cellcolor{blue!6}-0.09 & \cellcolor{red!8}0.13 & \cellcolor{red!10}0.17 & \cellcolor{red!34}0.57 & \cellcolor{red!60}1.00 & \cellcolor{blue!15}-0.25 & \cellcolor{blue!46}-0.77 \\
$a_\mu^{b,\rm th,LO}$ & \cellcolor{red!12}0.21 & \cellcolor{red!14}0.23 & \cellcolor{red!35}0.58 & \cellcolor{red!20}0.34 & \cellcolor{blue!47}-0.79 & \cellcolor{blue!15}-0.25 & \cellcolor{red!60}1.00 & \cellcolor{red!48}0.81 \\
$a_\mu^{b,\rm had,LO}$ & \cellcolor{red!10}0.17 & \cellcolor{red!11}0.19 & \cellcolor{red!16}0.27 & \cellcolor{red!6}0.10 & \cellcolor{blue!54}-0.90 & \cellcolor{blue!46}-0.77 & \cellcolor{red!48}0.81 & \cellcolor{red!60}1.00 \\
\hline
\end{tabular}}
\end{table*}

\section{Experimental Data for Charm}\label{App:charmdata}

Detailed scrutiny of the experimental data sources for the charm sector. Table \ref{tab:data} includes the experiments, with the year of publication, the energy range in GeV covered, the amount of data points, whether published data reports the systematic uncertainty splitting or not, and a detailed reference. The table contains 3 blocs, separated by the particular energy range covered, as sub-threshold, open-charm, and high-energy region play different roles in our analyses.

\begin{table*}[t]
\caption{Experimental data entering the continuum analysis for the charm sector.
\label{tab:data}}
\renewcommand{\arraystretch}{1.5}
\setlength{\tabcolsep}{8pt}
\begin{tabular}{|l|c|c|c|c|c|}
\hline
Experiment & Year & Energy range (GeV) & Data points & Syst. splitting & Reference \\
\hline
\multicolumn{6}{|c|}{\textit{Sub-threshold region}
  ($\sqrt{s} < 2M_D = 3.730$~GeV)} \\
\hline
BES       & 2000 & $2.6 - 5$       & 4/6   & no splitting given & \cite{BES:1999wbx} \\
BES       & 2002 & $3.73 - 4.80$   & 10/85 & splitting given    & \cite{BES:2001ckj} \\
BES       & 2006 & $3.650 - 3.872$ & 9/71  & splitting given    & \cite{BES:2006pcm,BES:2006dso} \\
BES       & 2009 & $2.60 - 3.65$   & 3     & no splitting given & \cite{BES:2009ejh} \\
BES       & 2022 & $2.23 - 3.67$   & 14    & splitting given    & \cite{BESIII:2021wib} \\
PLUTO     & 1982 & $3 - 31$        & 3/51  & no splitting given & \cite{Criegee:1981qx} \\
MARK~I    & 1981 & $2.6 - 7.8$     & 12/78 & splitting given    & \cite{Siegrist:1981zp} \\
KEDR      & 2019 & $1.84 - 3.72$   & 22    & full matrix given  & \cite{Anashin:2015woa,Anashin:2016hmv,KEDR:2018hhr} \\
\hline
\multicolumn{6}{|c|}{\textit{Open-charm threshold region}
  ($2M_D \leq \sqrt{s} \leq 4.8$~GeV)} \\
\hline
BES       & 2000 & $2.6 - 5$       & 1/6   & no splitting given & \cite{BES:1999wbx} \\
BES       & 2002 & $3.73 - 4.80$   & 75/85 & splitting given    & \cite{BES:2001ckj} \\
BES       & 2004 & $3.773$         & 1     & one point only     & \cite{BES:2004hbv} \\
BES       & 2006 & $3.650 - 3.872$ & 62/71 & splitting given    & \cite{BES:2006pcm,BES:2006dso} \\
CB        & 1986 & $3.87 - 4.5$    & 98    & splitting given    & \cite{Osterheld:1986hw} \\
CLEO      & 2008 & $3.97 - 4.26$   & 13    & no splitting given & \cite{CLEO:2008ojp} \\
PLUTO     & 1982 & $3 - 31$        & 30/51 & no splitting given & \cite{Criegee:1981qx} \\
MARK~I    & 1981 & $2.6 - 7.8$     & 21/78 & splitting given    & \cite{Siegrist:1981zp} \\
\hline
\multicolumn{6}{|c|}{\textit{High-energy continuum}
  ($\sqrt{s} > 4.8$~GeV)} \\
\hline
BES       & 2000 & $2.6 - 5$         & 1/6   & no splitting given & \cite{BES:1999wbx} \\
CB (Run~1)& 1990 & $5.20 - 7.00$     & 4     & splitting given    & \cite{Edwards:1990pc} \\
CB        & 1990 & $5.00 - 7.40$     & 11    & splitting given    & \cite{Edwards:1990pc} \\
CLEO      & 1997 & $10.52$           & 1     & one point only     & \cite{CLEO:1997eca} \\
CLEO      & 2007 & $6.964 - 10.538$  & 7     & splitting given    & \cite{CLEO:2007suf} \\
MD1       & 1996 & $7.2 - 10.34$     & 31    & splitting given    & \cite{Blinov:1993fw} \\
PLUTO     & 1982 & $3 - 31$          & 18/51 & no splitting given & \cite{Criegee:1981qx} \\
MARK~I    & 1981 & $2.6 - 7.8$       & 45/78 & splitting given    & \cite{Siegrist:1981zp} \\
\hline
\end{tabular}%
\end{table*}

\section{Moment equations}

\begin{figure*}[t]
\centering
\begin{subfigure}[b]{0.48\textwidth}
\centering
\includegraphics[width=\columnwidth]{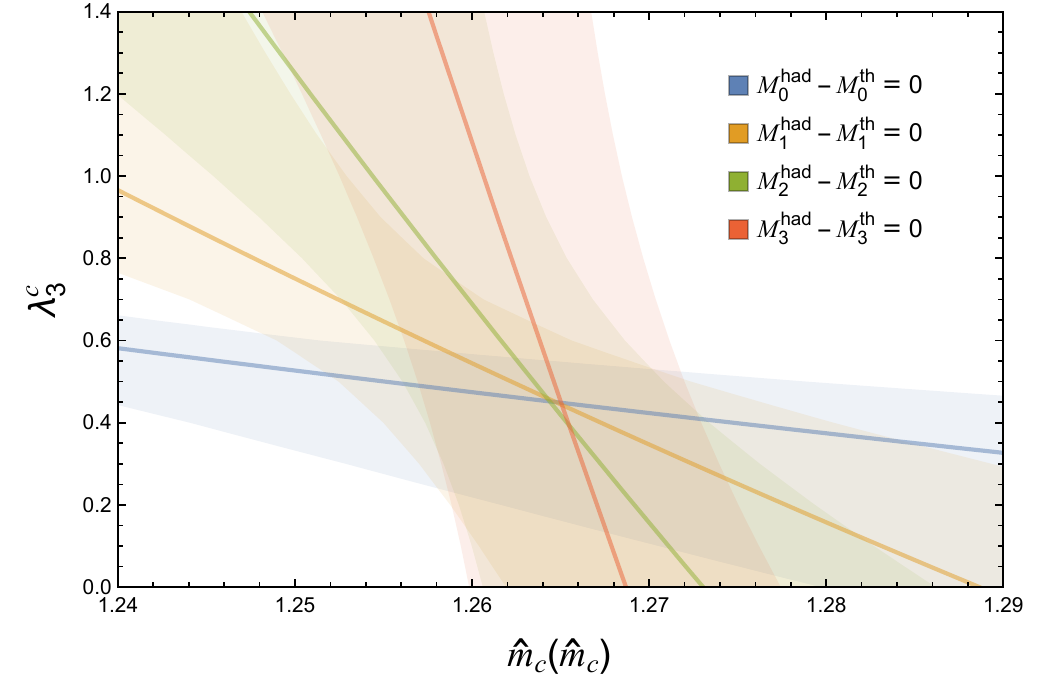}
\caption{}
\label{fig:bandplot_id}
\end{subfigure}
\hfill
\begin{subfigure}[b]{0.48\textwidth}
\centering
\includegraphics[width=\columnwidth]{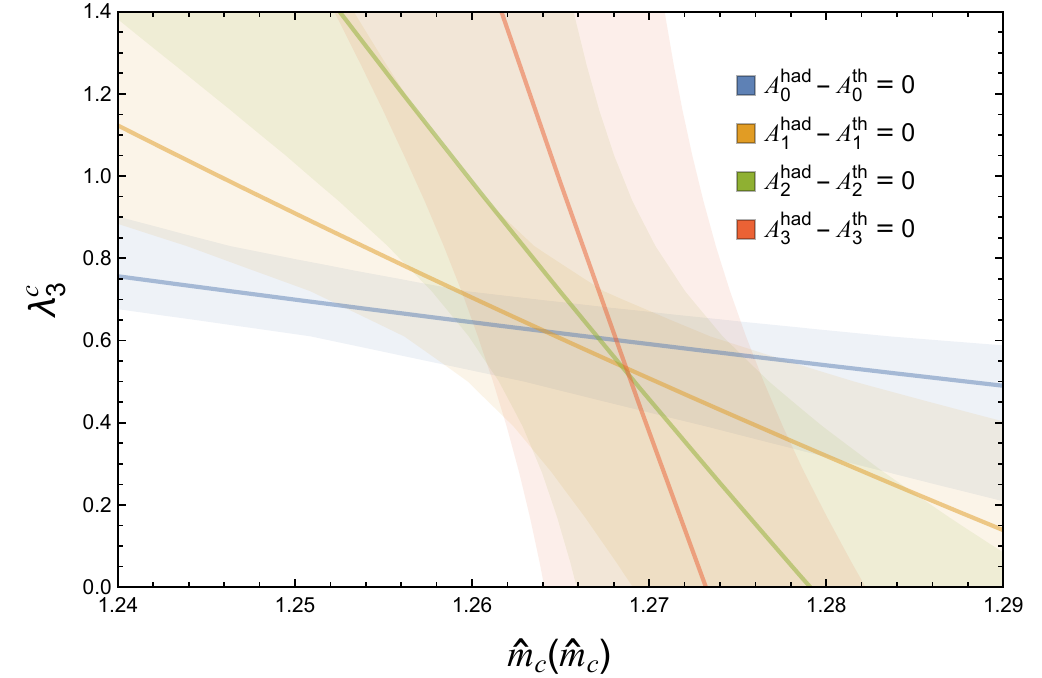}
\caption{}
\label{fig:bandplot_LO}
\end{subfigure}
\caption{Bands in the $(\hat m_c,\lambda_3^c)$ plane along which the consistency condition $\mathcal{A}_n^{\rm had}-\mathcal{A}_n^{\rm th}=0$ holds for a given moment $n$, including its uncertainty, for (a) the ordinary moments, $K=1$, and (b) the moments weighted by the kernel of the observable, $K=\hat K^{(2)}$. A pair of moments determines $(\hat m_c,\lambda_3^c)$ at the intersection of the corresponding two bands; the degree to which the remaining bands also pass through that intersection reflects the consistency of the extraction.}
\label{fig:bandplots_charm}
\end{figure*}

In this appendix, we illustrate in Figure \ref{fig:bandplots_charm} the role and correlation of the different moment sum rules for the extraction of the charm-quark mass and the $\lambda_3^c$ parameter, either using the unit kernel $K=1$ in Fig. \ref{fig:bandplot_id}, or the $K=\hat{K}^{(2)}$ in Fig. \ref{fig:bandplot_LO}. Our method makes use of two distinct sum rules to constrain the $(\hat m_c,\lambda_3^c)$ plane. The choice of the particular pair of moments is crucial for the result. These figures show the results of individual moments' sum rule where each moment equation is solved via a parameter scan of the variables $(\hat m_c,\lambda_3^c)$. Colored bands indicate the uncertainty propagation. The combination $0^{th}+3^{rd}$ is the one most perpendicular, thus introducing orthogonal information. On the other hand, the higher the moment, the larger the uncertainty, so orthogonality correlates with uncertainty. The $0^{th}+4^{th}$ pair is \textit{less} efficient as it induces larger uncertainties even though it may be more orthogonal than the $0^{th}+3^{rd}$. Either kernels yield similar results.

\end{document}